\documentclass[aps,preprint,groupedaddress,showpacs]{revtex4}

\begin{document}
\title{Nonlinear polarisation and dissipative correspondence between
	low frequency fluid and gyrofluid equations}
\author{Bruce D. Scott}
\email[email: ]{bds@ipp.mpg.de}
\homepage[\\ URL: ]{http://www.rzg.mpg.de/~bds/}
\affiliation{Max-Planck-Institut f\"ur Plasmaphysik, 
		Euratom Association,
                D-85748 Garching, Germany}

\date{\today}

\begin{abstract}
The correspondence between gyrofluid and low frequency fluid equations
is examined.  The lowest order conservative effects in ExB advection,
parallel dynamics, and curvature match trivially.  The principal
concerns are polarisation fluxes, and dissipative parallel viscosity and
parallel heat fluxes.  The emergence of the polarisation heat flux in
the fluid model and its contribution to the energy theorem is reviewed.
It is shown that gyroviscosity and the polarisation fluxes are matched
by the finite gyroradius corrections to advection in the long wavelength
limit, provided that the differences between gyrocenter and particle
representations is taken into account.  The dissipative parallel
viscosity is matched by the residual thermal anisotropy in the gyrofluid
model in the collision dominated limit.  The dissipative parallel heat
flux is matched by the gyrofluid parallel heat flux variables in the
collision dominated limit.  Hence, the gyrofluid equations are a
complete superset of the low frequency fluid equations.
\end{abstract}


\maketitle

\def\emskip{\hskip 1 em}
\def\hfb{\hfil\break}
\def\etc{{\it etc.}}
\def\visavis{{\it vis-a-vis}\ }
\def\ie{{\it i.e.}}
\def\eg{{\it e.g.}}
\def\etal{{\it et al}}
\def\ua{u.a.\ }
\def\dh{d.h.\ }
\def\zb{z.B.\ }
\def\bzw{bzw.\ }
\def\usw{usw.\ }

\def\idelta{$i$-delta}


\def\half{ {1\over 2} }
\def\third{ {1\over 3} }
\def\fourth{ {1\over 4} }
\def\tth{ {2\over 3} }
\def\twothirds{ {2\over 3} }
\def\threehalves{ {3\over 2} }
\def\fivehalves{ {5\over 2} }
\def\fivethirds{ {5\over 3} }
\def\sevenhalves{ {7\over 2} }
\def\threeh{\threehalves}
\def\eps{\epsilon}
 
\def\grapprox{\mathop{\lower.5ex \hbox{$\buildrel{\fivesy >}\over{\fivesy\sim}$}} \nolimits}
\def\lsapprox{\mathop{\lower.5ex \hbox{$\buildrel{\fivesy <}\over{\fivesy\sim}$}} \nolimits}
\def\grls{\mathop{\lower.5ex \hbox{$\buildrel{\fivesy >}\over{\fivesy <}$}} \nolimits}

\def\vec#1{{\bf #1}}
\def\tsr#1{{\secfnt #1}}
\def\avg#1{\left\langle #1 \right\rangle}
\def\abs#1{\left\vert #1 \right\vert}
\def\prf#1{\overline{#1}}

\def\max{{}_{{\rm max}}}
\def\min{{}_{{\rm min}}}

\def\minus{\mathop{\hbox{--}}\nolimits}

\def\re{\mathop{\rm Re}\nolimits}
\def\im{\mathop{\rm Im}\nolimits}
\def\sech{\mathop{\rm sech}\nolimits}
\def\diag{\mathop{\rm diag}\nolimits}
\def\Max{\mathop{\rm Max}\nolimits}
\def\Min{\mathop{\rm Min}\nolimits}
\def\nint{\mathop{\rm NINT}\nolimits}
\def\mod{\mathop{\rm mod}\nolimits}
\def\det{\mathop{\rm det}\nolimits}
\def\Tr{\mathop{\rm Tr}\nolimits}
\def\sign{\mathop{\rm sign}\nolimits}

\def\LBR{\left\lbrace}
\def\RBR{\right\rbrace}
\def\LB{\left\lbrack}
\def\RB{\right\rbrack}
\def\LP{\left (}
\def\RP{\right )}
\def\qq{\qquad\qquad}
\def\qqq{\qquad\qquad\qquad}
\def\Det#1{\left\vert\matrix{#1}\right\vert}

\def\pt{\partial}

\def\pzz#1{{\partial #1\over\partial z}}
\def\pxx#1{{\partial #1\over\partial x}}
\def\pyy#1{{\partial #1\over\partial y}}
\def\pww#1{{\partial #1\over\partial w}}
\def\pss#1{{\partial #1\over\partial s}}
\def\prr#1{{\partial #1\over\partial r}}
\def\prhrh#1{{\partial #1\over\partial \rho}}
\def\pthth#1{{\partial #1\over\partial \theta}}
\def\pchch#1{{\partial #1\over\partial \chi}}
\def\ppsps#1{{\partial #1\over\partial \psi}}
\def\pzeze#1{{\partial #1\over\partial \zeta}}
\def\pphph#1{{\partial #1\over\partial \phi}}
\def\ptt#1{{\partial #1\over\partial t}}
\def\pVV#1{{\partial #1\over\partial V}}
\def\phh#1{{\partial #1\over\partial \theta}}
\def\pvhvh#1{{\partial #1\over\partial \vartheta}}
\def\pxixi#1{{\partial #1\over\partial \xi}}
\def\dtt#1{{d #1\over dt}}
\def\dss#1{{d #1\over ds}}
\def\drr#1{{d #1\over dr}}
\def\pprr#1{{\partial^2 #1\over\partial r^2}}
\def\pprhrh#1{{\partial^2 #1\over\partial \rho^2}}
\def\ppss#1{{\partial^2 #1\over\partial s^2}}
\def\ppxx#1{{\partial^2 #1\over\partial x^2}}
\def\ppxy#1{{\partial^2 #1\over\partial x\partial y}}
\def\ppxs#1{{\partial^2 #1\over\partial x\partial s}}
\def\ppys#1{{\partial^2 #1\over\partial y\partial s}}
\def\ppyy#1{{\partial^2 #1\over\partial y^2}}
\def\ppzz#1{{\partial^2 #1\over\partial z^2}}
\def\pptt#1{{\partial^2 #1\over\partial t^2}}
\def\ppVV#1{{\partial^2 #1\over\partial V^2}}
\def\ppphph#1{{\partial^2 #1\over\partial \phi^2}}
\def\ppthth#1{{\partial^2 #1\over\partial \theta^2}}
\def\pphh#1{{\partial^2 #1\over\partial \theta^2}}
\def\ppvhvh#1{{\partial^2 #1\over\partial \vartheta^2}}
\def\ppxixi#1{{\partial^2 #1\over\partial \xi^2}}
\def\ppzeze#1{{\partial^2 #1\over\partial \zeta^2}}
\def\pphze#1{{\partial^2 #1\over\partial\theta\partial\zeta}}
\def\ppz#1{\partial #1/\partial z}
\def\ppx#1{\partial #1/\partial x}
\def\ppy#1{\partial #1/\partial y}
\def\ppw#1{\partial #1/\partial w}
\def\ppr#1{\partial #1/\partial r}
\def\pprh#1{\partial #1/\partial \rho}
\def\pps#1{\partial #1/\partial s}
\def\ppt#1{\partial #1/\partial t}
\def\ppV#1{\partial #1/\partial V}
\def\pph#1{\partial #1/\partial \theta}
\def\ppvh#1{\partial #1/\partial \vartheta}
\def\ppxi#1{\partial #1/\partial \xi}

\def\ddt#1{d #1/dt}
\def\pppz#1{\partial^2 #1/\partial z^2}
\def\pppx#1{\partial^2 #1/\partial x^2}
\def\pppy#1{\partial^2 #1/\partial y^2}
\def\pppr#1{\partial^2 #1/\partial r^2}
\def\ppprh#1{\partial^2 #1/\partial \rho^2}
\def\ppps#1{\partial^2 #1/\partial s^2}
\def\pppt#1{\partial^2 #1/\partial t^2}
\def\pppV#1{\partial^2 #1/\partial V^2}
\def\ppph#1{\partial^2 #1/\partial \theta^2}
\def\pppvh#1{\partial^2 #1/\partial \vartheta^2}
\def\pppxi#1{\partial^2 #1/\partial \xi^2}
\def\dddt#1{d^2 #1/dt^2}

\def\grad{\nabla}
\def\cross{{\bf \times}}
\def\div{\grad\cdot}
\def\divp{\grad_\perp\cdot}
\def\divpl{\grad_\parallel\cdot}
\def\curl{\grad\cross}
\def\dpl{\grad_\parallel}
\def\ddpl{\grad_\parallel^2}
\def\dpp{\grad_\perp}
\def\ddpp{\grad_\perp^2}
\def\delsq{\grad^2}
\def\delamb{ \mathchar"0274\hskip -.665em\mathchar"0275 }
\let\delam=\delamb
\def\lapl{\grad^2}
\def\lapldef{\ddpp=(\pt^2/\pt x^2)+K^2(\pt^2/\pt y^2)}

\def\pwww#1{{\partial #1\over\partial \vec w}}
\def\pwwpl#1{{\partial #1\over\partial w_\parallel}}
 
\def\pvv#1#2{{\partial #2\over\partial v_{#1}}}
\def\ppv#1#2{{\partial #2/\partial v_{#1}}}
\def\pvvv#1{{\partial #1\over\partial \vec v}}
\def\pvvp#1#2{{\partial #2\over\partial v'_{#1}}}
\def\ppvp#1#2{{\partial #2/\partial v'_{#1}}}
\def\pvvvp#1{{\partial #1\over\partial \vec v'}}
\def\pvvpl#1{{\partial #1\over\partial v_\parallel}}
 
\def\xunit{\vec{\hat x}}
\def\yunit{\vec{\hat y}}
\def\zunit{\vec{\hat z}}
\def\sunit{\vec{\hat s}}
\def\bunit{\vec{b}}
\def\eunit{\vec{\hat e}}
\def\nunit{\vec{\hat n}}
\def\dt{\Delta t}
\def\becomes{\leftarrow}
\def\from{\leftarrow}
\def\to{\rightarrow}
\def\fromto{\leftrightarrow}
\def\implies{\,\,\,\Longrightarrow\,\,\,}
\def\dotdot{\!:\!}

\def\meters{\,{\rm m}}
\def\invm{\,{\rm m}^{-3}}
\def\invmeter{\,{\rm m}^{-1}}
\def\invsec{\,{\rm sec}^{-1}}
\def\cm{\,{\rm cm}}
\def\km{\,{\rm km}}
\def\invcc{\,{\rm cm}^{-3}}
\def\invcm{\,{\rm cm}^{-1}}
\def\invmm{\,{\rm mm}^{-1}}
\def\mm{\,{\rm mm}}
\def\Vcm{\,{\rm V/cm}}
\def\Acm{\,{\rm A/cm^2}}
\def\kA{\,{\rm kA}}
\def\MA{\,{\rm MA}}
\def\degk{\,{\rm K}}
\def\ergs{\,{\rm erg}}
\def\eV{\,{\rm eV}}
\def\keV{\,{\rm keV}}
\def\MeV{\,{\rm MeV}}
\def\GeV{\,{\rm GeV}}
\def\kG{\,{\rm kG}}
\def\tesla{\,{\rm T}}
\def\kW{\,{\rm kW}}
\def\MW{\,{\rm MW}}
\def\MWsqm{\,{\rm MW/m^2}}
\def\Wsqm{\,{\rm W/m^2}}
\def\radsec{\,{\rm rad/sec}}
\def\Hz{\,{\rm Hz}}
\def\kHz{\,{\rm kHz}}
\def\MHz{\,{\rm MHz}}
\def\mpersec{\,{\rm m}/{\rm sec}}
\def\msqsec{\,{\rm m^2}/{\rm sec}}
\def\cmsec{\,{\rm cm}/{\rm sec}}
\def\kmsec{\,{\rm km}/{\rm sec}}
\def\mmsec{\,{\rm m}^2/{\rm sec}}
\def\msqsec{\,{\rm m}^2/{\rm sec}}
\def\cmcmsec{\,{\rm cm}^2/{\rm sec}}
\def\ccpersec{\,{\rm cm}^3/{\rm sec}}
\def\minutes{\,{\rm min}}
\def\yr{\,{\rm yr}}
\def\hr{\,{\rm hr}}
\def\Bar{\,{\rm bar}}
\def\sec{\,{\rm sec}}
\def\msec{\,{\rm msec}}
\def\usec{\,\mu{\rm sec}}

\def\ee{\vec E}
\def\bb{\vec B}
\def\ff{\vec F}
\def\jj{\vec J}
\def\qq{\vec q}
\def\aa{\vec A}
\def\kk{\vec k}
\def\vv{\vec v}
\def\uu{\vec u}
\def\xx{\vec x}
\def\ww{\vec w}

\def\bdel{\vec b\cdot\grad}
\def\Bdel{\vec B\cdot\grad}
\def\Jdel{\vec J\cdot\grad}
\def\bdot{\vec B\cdot}
\def\Bdot{\vec B\cdot}
\def\kdot{\vec k\cdot}
\def\exb{\vec E\cross\vec B}
\def\jxb{\vec J\cross\vec B}
\def\uxb{\vec u\cross\vec B}
\def\vxb{\vec v\cross\vec B}
\def\wxb{\vec w\cross\vec B}
\def\ucxb{{\vec u\over c}\cross\vec B}
\def\vcxb{{\vec v\over c}\cross\vec B}
\def\wcxb{{\vec w\over c}\cross\vec B}
\def\jcxb{{\vec J\cross\vec B\over c}}

\def\vexb{\vec v_E}
\def\vpol{\vec v_p}
\def\upol{\vec u_p}
\def\vstar{\vec v_*}
\def\ustar{\vec u_*}
\def\Jstar{\vec J_*}
\def\Jpol{\vec J_p}
\def\vgradb{\vec v_{\grad B}}
\def\qpol{\vec q_p}
\def\qstar{\vec q_\wedge}
\def\qestar{\vec q_e{}_\wedge}
\def\qistar{\vec q_i{}_\wedge}
\def\pistar{\vec\Pi_*}
\def\vR{\vec v_R}
\def\vdl{\vec v\cdot\grad}
\def\vdel{\vec v\cdot\grad}
\def\vedl{\vexb\cdot\grad}
\def\udl{\vec u\cdot\grad}
\def\udel{\vec u\cdot\grad}
\def\uidl{\vec u_i\cdot\grad}
\def\uidel{\vec u_i\cdot\grad}
\def\wdel{\vec w\cdot\grad}
\def\dedt#1{d_E #1/dt}
\def\dett#1{{d_E #1\over dt}}
\def\jpp{J_\perp}
\def\jperp{\vec\jpp}
\def\qpp{q_\perp}
\def\qperp{\vec\qpp}
\def\upp{u_\perp}
\def\uperp{\vec\upp}
\def\wpl{w_\parallel}
\def\wpp{w_\perp}
\def\wperp{\vec\wpp}
\def\vpp{v_\perp}
\def\vperp{\vec\vpp}
\def\lnb{\log B}
 
\def\rms{_{rms}}
 
\def\Jpl{J_\parallel}
\def\jpl{J_\parallel}
\def\Jpp{J_\perp}
\def\jpp{J_\perp}
\def\Jperp{\vec\Jpp}
\def\Bperp{\vec B_\perp}
\def\Apl{A_\parallel}
\def\apl{A_\parallel}
\def\App{A_\perp}
\def\app{A_\perp}
\def\Aperp{\vec\App}
\def\Epl{E_\parallel}
\def\epl{E_\parallel}
\def\Epp{E_\perp}
\def\epp{E_\perp}
\def\Eperp{\vec\Epp}
\def\upl{u_\parallel}
\def\vpl{v_\parallel}
\def\Upl{U_\parallel}
\def\vor{\grad_\perp^2\phi}
\def\kpl{k_\parallel}
\def\kkpl{k_\parallel^2}
\def\kpp{k_\perp}
\def\kperp{\vec\kpp}
\def\kkpp{k_\perp^2}
\def\xpl{{x_\parallel}}
\def\xpp{x_\perp}
\def\DD{\Delta_D}
\def\Dpl{D_\parallel}
\def\Dpp{\Delta_\perp}
\def\Depl{D_e{}_\parallel}
\def\Dipl{D_i{}_\parallel}
\def\Rpl{R_\parallel}
\def\qpl{q_\parallel}
\def\qepl{q_e{}_\parallel}
\def\qipl{q_i{}_\parallel}
\def\Pipl{\Pi_\parallel}
\def\qeperp{\vec q_e{}_\perp}
\def\qiperp{\vec q_i{}_\perp}
\def\mupl{\mu_\parallel}
\def\mupp{\mu_\perp}
\def\nuei{\nu_{ei}}
\def\nuee{\nu_{ee}}
\def\nuii{\nu_{ii}}
\def\wpe{\omega_{pe}}
\def\wpi{\omega_{pi}}
\def\nudamp{\nu_d}
\def\zeff{Z_{\!e\!f\!f}}
\def\lmfp{\lambda_{\!m\!f\!p}}
\def\ws{{\omega_*}}
\def\wsi{{\omega_{*i}}}
\def\wn{\omega_n}
\def\wt{\omega_t}
\def\wi{\omega_i}
\def\wT{\omega_T}
\def\wp{\omega_p}
\def\wc{{\omega_c}}
\def\kappacv{{\cal K}}
\def\wcv{{\omega_B}}
\def\etai{\eta_i}
\def\taui{\tau_i}
\def\rs{\rho_s}
\def\ld{\lambda_D}
\def\Lpl{L_\parallel}
\def\Lpp{L_\perp}
\def\lcorpl{\lambda_\parallel}
\def\lcorpp{\lambda_\perp}
\def\lcorx{\lambda_x}
\def\lcory{\lambda_y}
\def\rch{\rho_{ch}}
\def\npl{\eta_\parallel}
\def\etapl{\eta_\parallel}
\def\ald{a_L}
\def\alde{a_{Le}}
\def\aldi{a_{Li}}
\def\npp{\eta_\perp}
\def\etapp{\eta_\perp}
\def\kappapl{\kappa_\parallel}
\def\dprime{\Delta'}
\def\sk{{}_{\vec k}}
\def\sky{{}_{k_y}}
\def\gk{\gamma_k}
\def\vk{\vfl_k}
\def\nk{\nfl_k}
\def\tk{\tfl_k}
\def\dk{\Delta k}
\def\gd{\gamma_0}
\def\mwn{\Delta_n}
\def\mwh{\Delta_h}
\def\gamT{\Gamma_T}
\def\gamn{\Gamma_n}
\def\gamt{\Gamma_t}
\def\gami{\Gamma_i}
\def\gamc{\Gamma_c}
\def\gamk{\Gamma_k}
\def\gams{\Gamma_s}
\def\gaml{\Gamma_l}
\def\gamr{\Gamma_r}
 
\def\ptb{\widetilde}
\def\psifl{\widetilde\psi}
\def\phifl{\widetilde\phi}
\def\ffl{\widetilde f}
\def\fe{f_e}
\def\fefl{\widetilde f_e}
\def\fifl{\widetilde f_i}
\def\nfl{\widetilde n}
\def\hfl{\widetilde h}
\def\tfl{\widetilde T}
\def\nefl{\widetilde n_e}
\def\nifl{\widetilde n_i}
\def\tefl{\widetilde T_e}
\def\tifl{\widetilde T_i}
\def\pfl{\widetilde p}
\def\pefl{\widetilde p_e}
\def\pifl{\widetilde p_i}
\def\hefl{\widetilde h_e}
\def\vx{\widetilde v_x}
\def\vfl{\widetilde v}
\def\vefl{\widetilde \vexb}
\def\vxfl{\widetilde v_x}
\def\vyfl{\widetilde v_y}
\def\vrfl{\widetilde v_r}
\def\vppfl{\widetilde v_\perp}
\def\vflpp{\widetilde v_\perp}
\def\vplfl{\widetilde \vpl}
\def\Bfl{\widetilde \vec B}
\def\Bflpp{\widetilde B_\perp}
\def\Aplfl{\widetilde A_\parallel}
\def\Appfl{\widetilde A_\perp}
\def\Aperpfl{\widetilde {\vec A}_\perp}
\def\ufl{\widetilde u_\parallel}
\def\vorfl{\grad_\perp^2\phifl}
\def\jfl{\widetilde J_\parallel}
\def\qfl{\widetilde q_\parallel}
\def\qefl{\widetilde q_e{}_\parallel}
\def\qifl{\widetilde q_i{}_\parallel}
\def\jppfl{\widetilde J_\perp}
\def\jperpfl{\widetilde {\vec J}_\perp}
\def\Afl{\ptb A_\parallel}
\def\Jfl{\ptb J_\parallel}
\def\efl{\widetilde E_\parallel}
\def\Efl{\widetilde E_\parallel}
\def\Eppfl{\widetilde E_\perp}
\def\Eperpfl{\widetilde {\vec E}_\perp}
\def\etafl{\widetilde\eta}
\def\isatfl{\widetilde I_{{\rm sat}}}
\def\phiflfl{\widetilde\phi_{{\rm fl}}}
 
\def\teplfl{\widetilde T_e{}_\parallel}
\def\teppfl{\widetilde T_e{}_\perp}
\def\qeplfl{\widetilde q_e{}_\parallel}
\def\qeppfl{\widetilde q_e{}_\perp}
\def\tiplfl{\widetilde T_i{}_\parallel}
\def\tippfl{\widetilde T_i{}_\perp}
\def\qiplfl{\widetilde q_i{}_\parallel}
\def\qippfl{\widetilde q_i{}_\perp}

\def\tepl{ T_e{}_\parallel}
\def\tepp{ T_e{}_\perp}
\def\qepl{ q_e{}_\parallel}
\def\qepp{ q_e{}_\perp}
\def\tipl{ T_i{}_\parallel}
\def\tipp{ T_i{}_\perp}
\def\qipl{ q_i{}_\parallel}
\def\qipp{ q_i{}_\perp}

\def\peplfl{\widetilde p_e{}_\parallel}
\def\peppfl{\widetilde p_e{}_\perp}
\def\piplfl{\widetilde p_i{}_\parallel}
\def\pippfl{\widetilde p_i{}_\perp}

\def\pepl{ p_e{}_\parallel}
\def\pepp{ p_e{}_\perp}
\def\pipl{ p_i{}_\parallel}
\def\pipp{ p_i{}_\perp}


\def\phinn{ {e\phifl\over T} }
\def\nnn{ {\nfl\over n} }
\def\tnn{ {\tfl\over T} }
\def\unn{ {\ufl\over c_s} }
\def\vornn{ \rho_s^2\ddpp\phinn }
\def\jnn{ {\jfl\over ne} }
\def\qnn{ {\qfl\over nT} }
\def\psinn{ {\psifl\over B\rho_s} }

\def\ahat{\hat\alpha}
\def\ehat{\hat\eta}
\def\khat{\hat\kappa}
\def\shat{\hat s}
\def\bhat{\hat\beta}
\def\muhat{\hat\mu}
\def\epss{\hat\epsilon}
\def\bigpoint#1{
    \par\bigskip
    {\baselineskip=\normalbaselineskip
    \parindent=0 pt
    {\hfill\vbox{ #1  }\hfill}}
    \par\bigskip
    }
 
\def\jfm#1{{\it J. Fluid. Mech.} {\secfnt #1}}
\def\jgr#1{{\it J. Geophys. Res.} {\secfnt #1}}
\def\prl#1{{\it Phys. Rev. Lett.} {\secfnt #1}}
\def\physletta#1{{\it Phys. Lett. A} {\secfnt #1}}
\def\physlettb#1{{\it Phys. Lett. B} {\secfnt #1}}
\def\pf#1{{\it Phys. Fluids} {\secfnt #1}}
\def\pfa#1{{\it Phys. Fluids A} {\secfnt #1}}
\def\pfb#1{{\it Phys. Fluids B} {\secfnt #1}}
\def\physp#1{{\it Phys. Plasmas} {\secfnt #1}}
\def\nf#1{{\it Nucl. Fusion} {\secfnt #1}}
\def\njp#1{{\it New J. Phys.} {\secfnt #1}}
\def\cpp#1{{\it Contrib. Plasma Phys.} {\secfnt #1}}
\def\ppcf#1{{\it Plasma Phys. Contr. Fusion} {\secfnt #1}}
\def\plasphys#1{{\it Plasma Phys.} {\secfnt #1}}
\def\revpp#1{{\it Rev. Plasma Phys.} {\secfnt #1}}
\def\iaea#1#2{in {\it Plasma Physics and Controlled Nuclear Fusion
    Research #1}, Proceedings of the #2th International Conference}
\def\EPS#1#2#3{in {\it Proceedings of the
{#1}th European Conference on Controlled Fusion and Plasma Physics,
{#2}, {#3}} (European Physical Society, {#2}, {#3})}
\def\jcp#1{{\it J. Comput. Phys.} {\secfnt #1}}
\def\jetp#1{{\it Sov. Phys. JETP} {\secfnt #1}}
\def\sovjpp#1{{\it Sov. J. Plasma Phys.} {\secfnt #1}}
\def\jnm#1{{\it J. Nucl. Mat.} {\secfnt #1}}
\def\rsi#1{{\it Rev. Sci. Inst.} {\secfnt #1}}
\def\adv#1{{\it Adv. Phys.} {\secfnt #1}}
\def\apjl#1{{\it Astrophys. J. Lett.} {\secfnt #1}}
\def\apj#1{{\it Astrophys. J.} {\secfnt #1}}
\def\astrap#1{{\it Astron. Astrophys.} {\secfnt #1}}
\def\mnras#1{{\it MNRAS} {\secfnt #1}}
\def\vol#1{\ {\secfnt #1}}

\def\itemc{\hfb \null\hskip 20 pt $\circ$\ }
\def\itemcont{\\ \hskip 40 pt }

\def\qqquad{\qquad\qquad}

\def\sumsp{\sum_z}
\def\dV{dV\,}
\def\dW{dW\,}

\def\scripte{{\cal E}}
\def\drift{{c\over B^2}\vec B\cross}

\def\vor{\Omega}
\def\vorfl{\ptb\Omega}
\def\lcorx{\lambda_x}
\def\lcory{\lambda_y}

\def\bb{\vec B}
\def\uu{\vec u}
\def\fdot{\vec{\hat F}\cdot}
\def\fdl{\vec{\hat F}\cdot\grad}
\def\fdrxy{\hat F^{xy}}
\def\Bpp{B_\perp}
\def\vex{v_E^x}
\def\vey{v_E^y}
\def\bbx{b^x}
\def\bby{b^y}
\def\uex{u_E^x}
\def\uey{u_E^y}
\def\wex{w_E^x}
\def\wey{w_E^y}

\def\ppl{p_\parallel}
\def\ppp{p_\perp}
\def\deltap{\Delta p}
\def\deltat{\Delta T}

\def\ptb{\widetilde}

\def\vexfl{\ptb v_E^x}
\def\veyfl{\ptb v_E^y}
\def\bbxfl{\ptb b^x}
\def\bbyfl{\ptb b^y}
\def\uexfl{\ptb u_E^x}
\def\ueyfl{\ptb u_E^y}
\def\wexfl{\ptb w_E^x}
\def\weyfl{\ptb w_E^y}

\def\phig{\phi_G}
\def\vorg{\vor_G}
\def\phige{\phi_e}
\def\vorge{\vor_e}
\def\phigfl{\phifl_G}
\def\vorgfl{\vorfl_G}
\def\ufl{\ptb u_\parallel}
\def\Jfl{\ptb J_\parallel}

\def\nzfl{\ptb n_z}
\def\uzfl{\ptb u_z{}_\parallel}
\def\tzplfl{\ptb T_z{}_\parallel}
\def\tzppfl{\ptb T_z{}_\perp}
\def\qzplfl{\ptb q_z{}_\parallel}
\def\qzppfl{\ptb q_z{}_\perp}
\def\pzplfl{\ptb p_z{}_\parallel}
\def\pzppfl{\ptb p_z{}_\perp}
\def\Gzpl{G_z{}_\parallel}
\def\Gzpp{G_z{}_\perp}
\def\Wzpl{W_z{}_\parallel}
\def\Wzpp{W_z{}_\perp}
\def\Qzpl{Q_z{}_\parallel}
\def\Qzpp{Q_z{}_\perp}
\def\Gpl{G_\parallel}
\def\Gpp{G_\perp}
\def\Gplm{G_{m\parallel}}
\def\Gppm{G_{m\perp}}

\def\qeplpl{q_e{}_\parallel{}_\parallel}
\def\qepppl{q_e{}_\perp{}_\parallel}
\def\qiplpl{q_i{}_\parallel{}_\parallel}
\def\qipppl{q_i{}_\perp{}_\parallel}

\def\nefl{\ptb n_e}
\def\uefl{\ptb u_e{}_\parallel}
\def\teplfl{\ptb T_e{}_\parallel}
\def\teppfl{\ptb T_e{}_\perp}
\def\qeplfl{\ptb q_e{}_\parallel}
\def\qeppfl{\ptb q_e{}_\perp}
\def\peplfl{\ptb p_e{}_\parallel}
\def\peppfl{\ptb p_e{}_\perp}
\def\Gepl{G_e{}_\parallel}
\def\Gepp{G_e{}_\perp}
\def\Wepl{W_e{}_\parallel}
\def\Wepp{W_e{}_\perp}
\def\Qepl{Q_e{}_\parallel}
\def\Qepp{Q_e{}_\perp}

\def\nifl{\ptb n_i}
\def\uifl{\ptb u_i{}_\parallel}
\def\tiplfl{\ptb T_i{}_\parallel}
\def\tippfl{\ptb T_i{}_\perp}
\def\qiplfl{\ptb q_i{}_\parallel}
\def\qippfl{\ptb q_i{}_\perp}
\def\piplfl{\ptb p_i{}_\parallel}
\def\pippfl{\ptb p_i{}_\perp}
\def\Gipl{G_i{}_\parallel}
\def\Gipp{G_i{}_\perp}
\def\Wipl{W_i{}_\parallel}
\def\Wipp{W_i{}_\perp}
\def\Qipl{Q_i{}_\parallel}
\def\Qipp{Q_i{}_\perp}

\def\Nzfl{\LB n_z+\nzfl\RB}
\def\Tzplfl{\LB T_z+\tzplfl\RB}
\def\Tzppfl{\LB T_z+\tzppfl\RB}
\def\Pzplfl{\LB p_z+\pzplfl\RB}
\def\Pzppfl{\LB p_z+\pzppfl\RB}

\def\kkpp{k_\perp^2}

\def\uexb{\vec u_E}
\def\wexb{\vec w_E}
\def\Wexb{\vec W_E}

\def\uedl{\uexb\cdot\grad}
\def\wedl{\wexb\cdot\grad}
\def\Wedl{\Wexb\cdot\grad}

\def\pTT#1{{\pt #1\over\pt T}}

\def\bperp{\vec b_\perp}
\def\bbdl{\Bperp\cdot\grad}
\def\pxxmu#1{{\pt #1\over\pt x^\mu}}
\def\pxxnu#1{{\pt #1\over\pt x^\nu}}
\def\pyyk#1{{\pt #1\over\pt y_k}}
\def\ppyyk#1{{\pt^2 #1\over\pt y_k^2}}
\def\chiv{\hat\chi}

\def\kkpp{k_\perp^2}

\section{Introduction}

Low frequency reduced fluid equations are used to treat a variety of
phenomena in magnetised plasma dynamics, including turbulence
\cite{WakHas84,Waltz85} 
and tearing modes \cite{MonWhite80}.  The usual derivation path is to
start with the Braginskii collisional fluid equations \cite{Braginskii},
and then to solve for the velocity in the Lorentz force rather than the
inertia.  To lowest order in the inertia/gyrofrequency ratio, balance
among the principal forces is assumed, with pressure and the electric
forces balancing the magnetic force.  The result is a combination of
E-cross-B and diamagnetic flow terms, arising from the electric field
and pressure gradient, respectively.  In the conventional
magnetohydrodynamic limit
(MHD: considering a single velocity for all species and a single, total
pressure) the diamagnetic velocity is necessarily ordered small,
but in general the electron parallel dynamics
holds the electric and pressure forces to similar level.  The latter
is the ``adiabatic response'' which couples the electron pressure to the
electric parallel current through pressure forces and compressional
motion parallel to the background magnetic field.  Hence the MHD
ordering cannot be taken, and E-cross-B and diamagnetic flows are at
similar level.

The correction due to the inertia becomes the polarisation drift, which
is so called because it is opposite for electrons and ions.  The main
contribution in MHD ordering is due to the time dependence of the
electric field.  Nonlinear advection of the velocity field represents
the polarisation nonlinearity.  It is responsible for maintaining
drift wave self sustained turbulence \cite{ssdw,Camargo95,focus} and
also any Reynolds stress flow phenomena \cite{DiamondKim,sfdw}.  In
general the diamagnetic contributions enter at the same order and give
rise to what is generally called ``gyroviscosity'' --- the cancellation
of advection by the diamagnetic velocity in the equation of motion
\cite{HintonHorton71}. 
Once it is established that the polarisation drift enters at all, it is
necessary to keep it also in the ion temperature dynamics, since the
dynamics of the temperature and density are at similar order.  The
logical chain to this starts with the adiabatic response, including the
compression in the electron density equation, then noting the equality
of the electron and ion densities and the similar order in compression
of the polarisation and parallel currents, and finally the polarisation
drift entering the ion density and temperature at similar order
\cite{transport}.

Less familiar is the same phenomenon concerning the heat flux.  
The Braginskii model starts with a drifting Maxwellian distribution,
with not only arbitrary velocity but also velocity gradient,
to lowest order in the inertia/collision frequency ratio.  At next
order, velocity gradients appear but heat flux gradients do not.
This effectively and implicitly
assumes that heat fluxes are subthermal: the heat flux is assumed to be
smaller than the pressure times the velocity (see also Ref.\
\cite{Hassam80} for similar considerations regarding implicit
assumptions on the electron inertia and the magnetic current).
However, this is not true
even in the diamagnetic flows and heat fluxes, which in the presence of
temperature gradients are of similar strength.  The implicit assumption
of small heat fluxes breaks down completely.  When the MHD velocity
ordering also breaks down, it follows that the diamagnetic heat flux is
of similar magnitude as the pressure times the E-cross-B velocity.
This has been noted
before, by a treatment showing that the heat fluxes must be kept in the
gyroviscosity even to obtain the standard form of the polarisation
current \cite{Smolyakov97}.
However, one has to go further and consider inertia in the formulation
of the perpendicular heat flux itself \cite{Pogutse98}.
This is one order higher in the
moment hierarchy considered by Braginskii, which is why it is rarely
considered.  Nevertheless, polarisation enters the heat flux
equation as the correction due to finite inertia upon the diamagnetic
heat flux balance.  Then, since the polarisation enters the density and
temperature equations at the same order, and the polarisation heat flux
and velocities are also of the same order, the polarisation heat flux
should be considered in the temperature equation.  We will review this
herein as a preparation for establishing the correspondence between the
gyrofluid and low frequency Braginskii equations.  Ultimately,
correspondence is found in the nonlinear advection effects only if the
polarisation heat flux is kept in the fluid model.  

These polarisation phenomena enter the gyrofluid equations differently.
The gyrofluid equations have an entirely different derivation path
\cite{Dorland93,Beer96}, 
starting with the gyrokinetic equation with the low frequency and
small amplitude orderings already taken \cite{FriemanChen82}.
Polarisation enters the charge balance equation rather than the density
and temperature equations, 
since the latter are for the gyrocenters and not the particles
themselves.  The polarisation density balances differences in the
gyrocenter densities, maintaining quasineutrality \cite{Lee83}.  The
time derivative of this gyrokinetic polarisation equation gives a
relation analogous to the current balance (equivalently, vorticity)
equation in the fluid models, with the time derivative of the
polarisation density being the same as the divergence of the
polarisation current.  Underlying this is the Lie transformation between
particle and gyrocenter coordinates at the gyrokinetic level
\cite{Hahm88}.  Moments over this transform give the equations
describing the particle and gyrocenter representations of the moment
variables, corresponding to the fluid and gyrofluid models, respectively.

The gyrofluid equations are of significance because they allow treatment
of this drift dynamics at arbitrary order in the finite gyroradius
parameter (generally, the square of the perpendicular wavenumber
normalised to the gyroradius).  Tearing modes and reconnection involve
inertial layers which are thinner than the ion gyroradius
\cite{Coppi64}.  Tokamak edge
turbulence has a vorticity spectrum which always reaches down below the
ion gyroradius \cite{eps03}.  Treatment of these is generally beyond the
limits of equations whose derivation assumes the gyroradii are all
small.  Nevertheless, the low frequency Braginskii equations have a
systematic derivation, and it is desirable to know whether the gyrofluid
equations correspond properly to these under the limits within which the
Braginskii equations are perfectly valid.  That task is the purpose of
this work.  In the linear MHD limit the correspondence between
gyroviscosity and the finite gyroradius corrections in the ion density
equation were already shown \cite{Dorland93}.  Examination of the
nonlinear gyroviscous ``force'' in the MHD limit found certain
correspondences \cite{Belova01}.  Herein, we complete the correspondence
in the fully two-fluid limit.  It is recovered only if the polarisation
heat flux is kept in the fluid model.  Viewed another way, this effect
has always been present in the version of the gyrofluid model which
keeps perpendicular and parallel temperature moments
\cite{Dorland93,Beer96}.  The correspondence question is completed by
examining the dissipation model in the gyrofluid equations concerning
viscosity and parallel heat fluxes.  Ultimately, the gyrofluid equations
are found to recover the low frequency Braginskii equations, 
in the Braginskii limits of long
wavelengths, small heat fluxes, and complete collisional dominance.

The following sections respectively concern (II) 
the polarisation heat flux
and its effect on the free energy theorem
within the low frequency fluid equations, then (III) the correspondences
concerning polarisation in the density and temperature equations
including all the finite gyroradius nonlinearities, then (IV) the
collisional viscosity effects including correspondence to the
anisotropic corrections sometimes included in turbulence equations,
and also the contribution of heat fluxes to the viscosity, and then (V)
the parallel heat flux effects, whose correspondence is the easiest to
show.  The gyrofluid equations in question are from the most general GEM
(Gyrofluid ElectroMagnetic) model \cite{GEM}.  They will be introduced
piece by piece as needed.  The collisional fluid equations are much
better known --- see, e.g, the recent model including the anisotropy
effect in viscosity and the full polarisation velocity treatment in
Ref.\ \cite{Rogers98}.  A concluding commentary section (VI) is given at
the end.

\section{Polarisation including the heat flux}

Low frequency fluid equations can be derived directly using
the equation of motion for each species \cite{Braginskii},
\begin{equation}
nM\LP\ptt{\uu}+\udel\uu\RP + \div\Pi + \grad p
	= nZe\LP\ee+\ucxb\RP
\label{eqnativevel}
\end{equation}
solving for $\uu$ in the Lorentz force term to lowest order in
$\omega/\Omega_c\ll 1$ and assuming an electrostatic perpendicular
electric field with potential $\phi$ (justified by $\omega\ll\kpp v_A$;
i.e., the dynamics is too slow for dynamical Alfv\'enic compression),
\begin{equation}
\uperp^{(0)} = \drift\grad\phi + {1\over nZe}\drift\grad p
\end{equation}
noting this gives solely the perpendicular component.  The parallel
component has its own equation, derived separately.
The polarisation corrections are found by inserting this
$\uperp^{(0)}$ form into the inertia terms,
\begin{equation}
\uperp=\uperp^{(0)} + {M\over Ze}\drift\LP\ptt{\uperp^{(0)}}+\uperp^{(0)}\cdot\grad\uperp^{(0)}\RP 
	+ {1\over nZe}\drift\div\Pi(\uperp^{(0)})
\end{equation}
assuming flute mode ordering wherein $\upl\dpl\ll\uperp\cdot\grad$.
This is the standard version \cite{HintonHorton71}, usually behind the
derivation of the equations in turbulence models.  Alternatively, a
systematic procedure splitting the velocity into solenoidal and parallel
pieces \cite{Park84}, which can also include a potential-flow 
compressional piece \cite{Park87}, may be used.  The solenoidal flow
potential becomes $\phi$ under MHD ordering or generally a combination
of $\phi$ and $p$.  If drift ordering \cite{Rutherford68,Taylor68} is
then taken, the equations become identical to the reduced forms.
Drift ordering refers to the small amplitude but unity-order
nonlinearity limit used in the turbulence models.

Application of drift ordering to the velocity and including the
diamagnetic pieces in $\Pi$ results in the following form
\cite{HintonHorton71,Smolyakov97,Pogutse98},
\begin{equation}
\uperp = \drift\grad\phi + {1\over nZe}\drift\grad p
	- {Mc\over ZeB^2}\dtt{}\LP\grad\phi + {1\over nZe}\grad p\RP
\label{eqfluiddrift}
\end{equation}
where the $d/dt$ operator includes the nonlinear E-cross-B advection
\begin{equation}
\dtt{} = \ptt{} + \vedl \qqquad
\vexb = \drift\grad\phi
\end{equation}
The last term in Eq.\ (\ref{eqfluiddrift}) is the polarisation velocity,
whose charge-flux divergence is given by 
\begin{equation}
\div nZe\upol = -\div {nMc^2\over B^2}\dtt{}\LP\dpp\phi + {1\over nZe}\dpp p\RP
\end{equation}
In the conventional gyro-Bohm normalisation
for a single component plasma with singly charged ions this becomes
\begin{equation}
\div\upol = -\div \dtt{}\dpp(\phi + p_i) \equiv -\div \dtt{}\dpp W
\end{equation}
where $p_i=\tau_i(n_i + T_i)$ and $\tau_i$ is the background
ion/electron temperature ratio.  The pressure gradient is linearised,
and each species has its density and 
temperature normalised to its own background.
The flux and velocity divergence enter the same way because of the
normalisation and ordering.
The total ion force potential is denoted as $W$.
The time scale inferred by the divergence of the velocity is normalised
to the profile scale and the sound speed, $\Lpp/c_s$, where
$c_s^2=T_e/M_i$ and $L_{Te}$ is usually used for $\Lpp$.
The double perpendicular derivative is normalised to the square of the
drift scale, $\rs^2=c^2T_eM_i/e^2B^2$, and it is useful to note that
$\tau_i\rs^2=\rho_i^2$, which makes the role of the finite gyroradius
explicit.  The above considerations constitute what is also called local
ordering in the context of turbulence computation.

A similar treatment for the heat flux starts with the $(Mv^2/2)\vv$
moment of the kinetic equation, analogous to the $M\vv$ moment and the
equation of motion.  The heat flux equation is given by
\begin{equation}
\dtt{\vec q} + \fivehalves{p\over M}\grad T = {e\over Mc}\vec q\cross\vec B
\end{equation}
under drift ordering (Eq.\ 11 of Ref.\ \cite{Pogutse98}, after the
diamagnetic cancellation is taken),
Solving this to lowest order
neglecting the inertial effects yields the diamagnetic heat flux,
\begin{equation}
\qperp^{(0)} = \fivehalves {p\over Ze}\drift\grad T
\end{equation}
Using the same ordering and normalisation
conventions as for the velocity, we find the
divergence of the polarisation correction for singly charged ions,
\begin{equation}
\div\qpol = -\fivehalves\tau_i\div \dtt{}\dpp T_i
\end{equation}
under gyro-Bohm normalisation ($\vec q$ is normalised the same way as
$p\uu$).  Clearly, this is the same order as the 
velocity polarisation divergence if the gradients of the state
variables (potential, densities and temperatures) are all comparable.
The polarisation velocity divergence enters both the density and
temperature equations, and the polarisation heat flux divergence enters
the temperature equation.  
This sort of consistency is well known for the diamagnetic fluxes
themselves, since with similar $(e/T)\grad\phi$ and
$\grad\log T$ and $\grad\log n$ they all enter at the same order with
each other and with toroidal compression of the E-cross-B velocity
\cite{Weiland89,Weiland90} (cf.\ discussion and manipulations in Ref.\
\cite{Weiland}). 
but the same results concerning the polarisation fluxes was not widely
known before Refs.\ \cite{Smolyakov97,Pogutse98},
and is still routinely missed by low frequency fluid models.  

Under the above considerations we have the normalised ion density and
temperature equations with polarisation divergences,
\begin{eqnarray}
& &\dtt{n_i}-\div\dtt{}\grad W + \dpl\upl = \kappacv(W+G)
\\
& &\threehalves\dtt{T_i}-\div\dtt{}\grad W 
	-\fivehalves\tau_i\div\dtt{}\grad T_i + \dpl(\upl+\qipl) 
	= \kappacv(W+G)+\fivehalves\tau_i\kappacv(T_i)
\end{eqnarray}
The parallel velocity divergence is included in both equations and the
heat flux divergence in the temperature equation.  The terms denoted by
$\kappacv$ are the remnant divergences of the E-cross-B and diamagnetic
velocities (represented in total by $W$) and diamagnetic specific heat
flux (represented by $5\tau_iT_i/2$), due to the inhomogeneous magnetic
field, after the diamagnetic cancellation is taken in the temperature
equation \cite{Tsai70}.  The curvature operator is then defined, e.g.,
\begin{equation}
\kappacv(\phi) = -\div\vexb = -\div\drift\grad\phi
\end{equation}
in terms of the E-cross-B divergence.
The quantity $G$ in the ion density and temperature equations arises
from thermal anisotropy.  It is given by \cite{Rogers98},
\begin{equation}
G = {0.96\over 12\nu_i}\LB\kappacv(W) - 4\dpl\upl\RB
\end{equation}
in the collisional limit and represents viscous dissipation, with
$\nu_i$ the ion collision frequency normalised to $c_s/\Lpp$.

For the electrons
the convention is to neglect the mass everywhere except in parallel
inertia (entering the parallel velocity and heat flux equations).
The electron density equation is given by
\begin{equation}
\dtt{n_e} + \dpl\vpl = \kappacv(\phi-p_e)
\end{equation}
in which anisotropy and polarisation (electron viscosity and inertia)
are neglected.  The electron pressure gradient is linearised in the same
was as for the ions, with the minus sign reflecting the normalised
temperature/charge ratio.
The quasineutrality condition is given by the
subtraction of the two density equations and neglecting the space charge
density, so that
\begin{equation}
\div\dtt{}\grad W = \dpl(\upl-\vpl) - \kappacv(p_e+p_i+G)
\end{equation}
equivalently, $\div\jj=0$, whose three pieces are the polarisation,
parallel, and diamagnetic divergences, respectively.  We note that
\begin{equation}
\Jpl = \upl-\vpl
\end{equation}
defines the parallel current (under the normalisation);
this is usually used to eliminate $\vpl$ in favour of $\Jpl$.
We may further
subtract this from the ion temperature equation to obtain
\begin{equation}
\threehalves\dtt{T_i}-\fivehalves\tau_i\div\dtt{}\grad T_i 
	+ \dpl(\vpl+\qipl) = \kappacv(\phi-p_e)
	+\fivehalves\tau_i\kappacv(T_i)
\end{equation}
eliminating the polarisation divergence in the velocity but not the heat
flux.  In this equation, the explicit ion velocity divergences are
replaced by the electron ones, but the ion heat flux divergences
remain.  These are the polarisation and diamagnetic heat flux terms,
respectively the second and last terms in the line above.

\subsection{Free energy in the fluid model
\label{secfluidenergy}}

The complete set of equations in the fluid model is given by
\begin{equation}
\div\dtt{}\grad W = \dpl\Jpl - \kappacv(p_e+p_i+G)
\label{eqvor}
\end{equation}
\begin{equation}
\dtt{n_e} + \dpl\vpl = \kappacv(\phi-p_e)
\label{eqne}
\end{equation}
\begin{equation}
\threehalves\dtt{T_e} + \dpl(\vpl+\qepl) 
	= \kappacv(\phi-p_e)-\fivehalves\kappacv(T_e)
\label{eqte}
\end{equation}
\begin{equation}
\threehalves\dtt{T_i}-\fivehalves\tau_i\div\dtt{}\grad T_i 
	+ \dpl(\vpl+\qipl) = \kappacv(\phi-p_e)
	+\fivehalves\tau_i\kappacv(T_i)
\label{eqti}
\end{equation}
\begin{equation}
\dtt{\upl} + \dpl(p_e+p_i+4G) = 0
\label{equpl}
\end{equation}
\begin{equation}
\beta_e\ptt{\Apl}+\mu_e\dtt{\Jpl} + \dpl(\phi-p_e) = - R_{ei}
\label{eqjpl}
\end{equation}
where $n_i$ and $n_e$ are equivalent, $\vpl$ is given by $\upl-\Jpl$,
and the parallel heat fluxes
$\qepl$ and $\qipl$ and the resistive dissipation $R_{ei}$
are
left undetermined (at this level they may be given their Braginskii
dissipative formulae \cite{Braginskii}).  The factor of $4G$ in the
parallel momentum equation is also the result of anisotropy.  
Except for the retentions
of the polarisation heat flux in Eq.\ (\ref{eqti})
and the electron inertia in
Eq.\ (\ref{eqjpl}), these equations are the same as those
given in Ref.\ \cite{Rogers98}.  The normalisation convention is the
standard gyro-Bohm one, with $\dpl$ normalised against $\Lpp$, not 
$qR$, which is why the un-scaled forms for 
\begin{equation}
\beta_e={4\pi p_e\over B^2} \qqquad \mu_e={m_e\over M_D}
\end{equation}
are used.  The factor $\tau_i$ gives the background $T_i/T_e$ ratio.
The pressures are linearised as above.

The free energy of the system is given by
\begin{eqnarray}
& &\scripte = \int\dV\half\LB \abs{\dpp W}^2 + (1+\tau_i)n_e^2
	+ \threehalves T_e^2 
	+ \threehalves \tau_i T_i^2
\right.\nonumber\\ & & \left.\hskip 2 cm{}
	+ \upl^2
	+ \beta_e\abs{\dpp\Apl}^2 + \mu_e\Jpl^2
	+ \fivehalves\abs{\tau_i\dpp T_i}^2\RB
\label{eqenergy}
\end{eqnarray}
where $\int\dV$ denotes complete spatial integration and
now the $\tau_i$ factors are put in explicitly.
Except for the last term, due to the polarisation heat flux, this has
been analysed before \cite{pet97,hmode97}.  
Insertion of Eqs.\ (\ref{eqvor}-\ref{eqjpl})
into $\ppt{\scripte}$ finds this time derivative to vanish except for
the dissipative terms (here, there are no gradient source terms since
the profile gradients are kept within the dependent variables of the
model; without explicit sources this corresponds to decaying cases
initialised with a finite profile and a random bath of fluctuations, as
in Ref.\ \cite{gem3test}).

The last term in Eq.\ (\ref{eqenergy}) represents the polarisation heat flux.
If Eq.\ (\ref{eqti}) is multiplied by $\tau_i T_i$ and integrated, the
$\ppt{}$ terms yield the two terms in Eq.\ (\ref{eqenergy})
explicitly dependent upon $T_i$.
If Eq.\ (\ref{eqvor}) is multiplied by $W$ and integrated, the
$\ppt{}$ term yields the term in Eq.\ (\ref{eqenergy})
explicitly dependent upon $W$.  The resulting term $W\dpl\Jpl$ is
balanced by the contributions $\tau_i n_e$ and $\tau_i T_i$ times
$\dpl\Jpl$ in Eqs.\ (\ref{eqne},\ref{eqti}) and the contribution
$\Jpl\dpl\phi$ coming from Eq.\ (\ref{eqjpl}).
And henceforth.  
The appearance of a heat flux term in the energy may be unfamiliar in a
fluid model, but it is known from the gyrofluid model (cf.\ Ref.\
\cite{Sugama01}\ and below) and also from treatments of extended fluid
dynamics \cite{thermalwaves}.  Due to the relation between the
polarisation heat flux (as, up to coefficients, the time derivative of
the curl of) the diamagnetic heat flux, the energy contribution 
is equivalent to $2/5$ times the square of the diamagnetic heat flux (up
to normalisation).  This is equivalent to the relation between the
polarisation velocity and the E-cross-B and diamagnetic velocities, and
the appearance of the square of the latter in the form of the
perpendicular kinetic energy.  For general gradient driven turbulence,
the heat flux and velocity pieces are of comparable magnitude.

The rest of the fluid model is given by the dissipation due to parallel
heat fluxes and viscosity.  The latter comes from temperature
anisotropy, in this case $G$.  Starting with a diagonal pressure tensor
\begin{equation}
\vec P = \diag\{p_\perp,p_\perp,p_\parallel\}
\end{equation}
we split it into isotropic and traceless parts,
\begin{equation}
\vec P = p\vec g + \Pi \qqquad
\Pi = 2G\diag\{-1,-1,2\}
\end{equation}
where $\vec g$ is the metric tensor, identifying $6G$ with the
anisotropy $\deltap=p_\parallel-p_\perp$ (see Section IV, below). 
With the factor of
density common, $\deltap$ is equivalent to $\deltat$.  The viscosity
model sets collisional dissipation of $G$ against perpendicular and parallel
velocity divergences as above.  Similarly, the parallel heat flux
formulae are given by setting their collisional dissipation against the
corresponding parallel temperature gradients, e.g., 
\begin{equation}
{5/2\over\kappa_i}\nu_i\qipl = -\fivehalves\tau_i\dpl T_i
\end{equation}
for ions, with the coefficient set such that the familiar formula
\cite{Braginskii}\ with $\kappa_i=3.9$ results.

The remainder of this paper is concerned with recovery of these formulae
(polarisation effects in the vorticity and ion temperature equations,
viscosity through anisotropy, and the parallel heat fluxes), from the
gyrofluid model under the same ordering conventions as for this one.

\subsection{Interlude --- bracket notation
\label{secbrackets}}

In several treatments of the equations of turbulence in confined plasmas
the nonlinearities are explicitly written in a form which makes their
conservation properties obvious.  Basically, e.g., $\vedl n_e$ is
written as $[\phi,n_e]$, where the bracket involves the perpendicular
derivatives involved in the drift motion.  It is variously written as
\begin{equation}
[\phi,n_e] = \pxx{\phi}\pyy{n_e} - \pxx{n_e}\pyy{\phi}
\label{eqslab}
\end{equation}
in slab or local fluxtube treatments, or as
\begin{equation}
[\phi,n_e] = {1\over r}\LP\prr{\phi}\phh{n_e} - \prr{n_e}\phh{\phi}\RP
\label{eqcylinder}
\end{equation}
in ``cylinder'' treatments, which with the field aligning coordinate
transformations $x=r^2/a^2$ and $y_k=q(\theta-\theta_k)-\zeta$ and $s=\theta$
with $q=q(r)$ becomes
\begin{equation}
[\phi,n_e] = {2\over a^2}\LB\pxx{\phi}\LP q\pyy{n_e}+\pss{n_e}\RP
	- \pxx{n_e}\LP q\pyy{\phi}+\pss{\phi}\RP\RB
\label{eqgemr}
\end{equation}
at $s=\theta_k$ since $\ppr{y_k}$ vanishes there.  Under fluxtube
ordering $\pps{}$ is small compared to either $\ppx{}$ or $\ppy{}$ and
the factor of $q$ is replaced by a constant, 
and Eq.\ (\ref{eqgemr}) reverts to the form in Eq.\ (\ref{eqslab})
with a multiplier of $2q/a^2$ which can be normalised away.
More detail on this is given in Ref.\ \cite{shifted}.

With the bracket notation we make use of the following properties
\begin{equation}
\int\dV [f,g] = 0 \qqquad f[f,g] = \half[f^2,g]
	\qqquad g[f,g] = \half[f,g^2]
\end{equation}
in the manipulations below.
That is, the bracket is a perfect divergence, and both energy and
entropy are conserved.  Useful manipulations include
\begin{equation}
\ddpp[f,g] = \div [f,\dpp g] + \div [\dpp f,g]
\end{equation}
\begin{equation}
\div [f,\dpp g] = [\dpp f,\dpp g] + [f,\ddpp g]
\end{equation}
In any of these the perpendicular subscript may be regarded as understood, as is
the contraction implied by $[\grad f,\grad g]$.

\subsection{Free energy, adiabatic response, and MHD ordering}

The fluid model's vorticity equation in this notation is
\begin{equation}
\dtt{}\ddpp W + [\dpp\phi,\dpp W] = \dpl\Jpl - \kappacv(p_e+p_i+G)
\label{eqfluidvor}
\end{equation}
where use is made of 
\begin{equation}
\dtt{}=\ptt{}+[\phi,]
\end{equation}
Since the bracket is antisymmetric, the second term in Eq.\
(\ref{eqfluidvor}) is equivalent to $[\dpp\phi,\dpp p_i]$, and hence
this ``gyroviscous correction'' is a proper warm-ion effect.  The
polarisation terms may be manipulated to show 
\begin{equation}
\ptt{}\ddpp W + \div[\phi,\dpp W] = 
	\ptt{}\ddpp \phi + \div[W,\dpp \phi] + \ddpp\dtt{p_i}
\end{equation}
Under MHD ordering the last term on the right hand side
is dropped, as was done in Ref.\
\cite{HintonHorton71}, unfortunately without the explicit statement that
this depends strictly on $\dpp p_i\ll n_e e\dpp\phi$ remaining valid
(un-normalised units).

MHD ordering implies the lack of a $\tau_i\dpl\Jpl$ term in the
energetics.  This can only be reconciled if $\dpl\Jpl$ is neglected in
all of the continuity equations, not just the one for $n_e$,
since quasineutrality ties the ion and electron dynamics together.  If
$\dpl\Jpl$ is neglected then $\dpl p_e$ itself must be neglected in 
Eq.\ (\ref{eqjpl}).  Hence, we would be back not only to MHD ordering
but to reduced MHD itself.  This may be demonstrated by alternatively
multiplying Eq.\ (\ref{eqvor}) by $\phi$ or $W$ and integrating, and
observing the logical consequences.  Either reduced MHD is taken in its
entirety, or the consequences of the adiabatic response are taken to
their conclusion.  No intermediate version is energetically closed
\cite{transport}.

It is the adiabatic response in the electron dynamics which disallows
the use of MHD ordering in this kind of turbulence.  In Eq.\
(\ref{eqjpl}) the two static force terms (those not dependent on $\Jpl$
or $\Apl$) are often the largest, and even for edge turbulence there is
a partial cancellation between them.  Their difference determines
$\Jpl$, mediated by induction, inertia, or resistivity according to
whether finite $\beta_e$, finite $\mu_e$, or finite $\nu_e$ is the
strongest.  Once $\dpl p_e$ is kept in Eq.\ (\ref{eqjpl}) then
$\dpl\Jpl$ must be kept in Eq.\ (\ref{eqne}), since in the energetics
$n_e\dpl\Jpl+\Jpl\dpl n_e$ must become a total divergence.  Since it is
$p_e$ which appears in Eq.\ (\ref{eqjpl}), then $\dpl\Jpl$ must also be
kept in Eq.\ (\ref{eqte}).  Now, the densities $n_e$ and $n_i$ are
equivalent, so the appearance of $\dpl\Jpl$ in Eq.\ (\ref{eqne}) implies
the appearance of $\div\upol$ in the equation for $n_i$, which is made
explicit if Eqs.\ (\ref{eqne},\ref{eqvor}) are subtracted (eliminating
the $\dpl\Jpl$ term).  To conserve against the implied $\tau_i\dpl\Jpl$
term, the pressure effects in polarisation must be kept since Eq.\
(\ref{eqvor}) must be multiplied by $W$ rather than $\phi$, since $W$
contains $\tau_i n_e$.  However, $W$ also contains $\tau_i T_i$.  Hence,
$\div\upol$ must also be kept in Eq.\ (\ref{eqti}), wherein it has been
replaced by $\dpl\Jpl$.  Only now is the energetic loop started by $\dpl
p_e$ in Eq.\ (\ref{eqjpl}) closed, since now $-W\dpl\Jpl$ closes against
$\tau_i(n_i+T_i)\dpl\Jpl$ from Eqs.\ (\ref{eqne},\ref{eqti}) as well as
against $-\Jpl\dpl\phi$ from Eq.\ (\ref{eqjpl}).  Hence the logical
chain: adiabatic response in the equation for $\Jpl$, parallel
compression in the equations for $n_e$ and $T_e$, quasineutrality, by
which $\dpl\Jpl$ implies $\div\upol$, and finally the qualitative
similarity among densities and temperatures.
All this is forced by the similarity in magnitude among
$\dpp\{\phi,n_e,T_e,T_i\}$, caused by the adiabatic response.
Similar consequences may be found among the curvature terms ---
essentially, the retention of diamagnetic compression in the continuity
equations also forces the complete two fluid version of ion polarisation.
The above analysis has been given for the local form of the equations;
the corresponding one for the global form was given in Ref.\
\cite{transport}. 

\subsection{Interlude --- relation to extended fluid and extended MHD models}

The low frequency fluid models are themselves distinct from what is
called extended fluid dynamics or extended MHD \cite{Ramos05,Jardin05,
Catto04,Simakov03, Sovinec03,Sugiyama00,Park99,ChangCallen92}.  In the
latter the low frequency is invoked to obtain expressions for heat
fluxes and viscosities.  These are not, however, expressed in terms of
drifts and polarisation, but left in the native form with velocity and
heat flux vectors.  Specifically, the steps in Eqs.\
(\ref{eqnativevel}--\ref{eqfluiddrift}) are not taken.  Moreover,
explicit time dependence of the heat flux and dissipative viscosities
are neglected (thermal anisotropy is assumed to be small -- see below).
Fluid drift theory replaces the vector forms with scalar quantities in
the list of dependent variables; for example, $\phi$ and $p$ and $\upl$
are the variables with which $\vec u$ is described.  Further to that are
the gyrofluid models \cite{Dorland93,Beer96,aps99,GEM}, to be discussed
below.  No matter the complexity, extended fluid models and fluid drift
models break down when $\kpp\rho_i$ becomes unity or larger, as it
always does in tokamak edge turbulence \cite{aps99,eps03,eps06,krakow}.
Gyrofluid models are required to overcome this.  Of course, since they
have a different formulation, it is desired to know how well they
recover the fluid forms when the latter are valid.  That is the point of
this work.

\section{Gyrofluid FLR nonlinearities and fluid gyroviscosity}

The gyrofluid model has a different structure from the fluid one ---
like the underlying gyrokinetic model, the moment variables (the model
for the kinetic distribution function) are advanced independently for
each species, and then the field equations (polarisation and induction)
are solved for the electrostatic and parallel magnetic potentials.
Analysis of the gyrofluid moment equations in the various limits
proceeds the same way for each species.   Due to the correspondences
involved we concentrate mainly on the ions.  Consideration of thermal
forces at the end will then involve the electrons.  This section is
concerned with the gyroviscosity effects in the fluid model, meaning
essentially all the differences in the polarisation between the general
one and the MHD one (the latter involving $\phi$ only).  We will show
how these emerge naturally from the finite gyroradius (FLR)
nonlinearities in the ion gyrocenter density and temperature equations
in the limit of small $\kpp\rho_i\ll 1$.

The equations under consideration are for the density and the parallel
and perpendicular temperatures (Eqs.\ 99,101,102 of Ref.\ \cite{GEM}),
\begin{equation}
\ptt{n_i} + [\phig,n_i] + [\vorg,\tipp]
	+ \dpl\upl = \kappacv\LP\phig+{\pipl+\pipp+\vorg\over 2}\RP
\label{eqni}
\end{equation}
\begin{equation}
\half\ptt{\tipl} + \half[\phig,\tipl] 
	+ \dpl(\upl+\qiplpl) 
	= \kappacv\LP{\phig+\pipl+2\taui\tipl\over 2}\RP
	- 2\nu_i G
\label{eqtipl}
\end{equation}
\begin{eqnarray}
& &\ptt{\tipp} + [\phig,\tipp] + [\vorg,n_i+2\tipp]
	+ \dpl\qipppl 
	\nonumber\\ & &\qquad{}
	= \half\kappacv\LP{\phig+4\vorg+\pipp+3\taui\tipp\over 2}\RP
	+ 2\nu_i G
\label{eqtipp}
\end{eqnarray}
where in terms of the perpendicular and parallel pressures the isotropic one is
$p_i=(\pipl+2\pipp)/3$ and the difference is
$G=(p_\parallel-p_\perp)/6$, dissipated in the term proportional to
$\nu_i G$.  Normalisation is to a common background temperature $T_0$,
with $\taui=T_i/ZT_0$ giving the temperature/charge ratio.  The factor
of $\taui$ is folded into the pressures.
The heat fluxes are also broken up into
parallel transport of perpendicular and parallel energy, the $(M\wpl^2/2)\wpl$
and $(M\wpp^2/2)\wpl$ moments, respectively, where $\vec w$ is the
kinetic velocity in the co-moving reference frame.  In the gyrokinetic
and gyrofluid models $\vec w_\perp$ is not used directly, but as
$\wpp^2$, specifically, the magnetic moment $\mu=M\wpp^2/2B$, due to the
low frequency ordering.  These heat flux pieces are
$\qiplpl$ and $\qipppl$, respectively.  Finally, the potentials $\phig$
and $\vorg$ represent the FLR treatment.  The Pad\'e approximants are
\begin{equation}
\phig = {\phi\over 1+b/2} \qqquad \vorg = {-b^2\phi/2\over (1+b/2)^2}
\label{eqphig}
\end{equation}
in wavenumber space, with argument $b=\kkpp\rho_i^2$.  In the limit of
$\kkpp\to 0$ we have
\begin{equation}
-b\to \tau_i\ddpp \qqquad \phig\to (1-b/2)\phi \qqquad \vorg\to (-b/2)\phi 
\end{equation}
which we will use to show correspondence.  The notation of Ref.\
\cite{GEM} is used, and the FLR treatment follows Refs.\
\cite{Dorland93,Beer96}\ with the necessary modifications to restore
free energy conservation as discussed in Ref.\ \cite{GEM}.

\subsection{The isothermal version
\label{secisothermal}}

To make the analysis easier to follow, we start with the isothermal
model which neglects all considerations of temperature dynamics,
including the anisotropy and heat fluxes.  In this case the ion and
electron density equations are given by
\begin{equation}
\ptt{n_i} + [\phig,n_i] + \dpl\upl = \kappacv(\phig+\tau_i n_i)
\label{eqnia}
\end{equation}
\begin{equation}
\ptt{n_e} + [\phi,n_e] + \dpl\vpl = \kappacv(\phi-n_e)
\label{eqnea}
\end{equation}
and they are related through the polarisation equation,
\begin{equation}
{n_i\over 1+b/2} + \tau_i^{-1}{b\over 1+b}\phi = n_e
\label{eqpola}
\end{equation}
using the Pad\'e approximants (cf.\ Refs.\ \cite{Beer96,GEM}).
This determines the gyrocenter density $n_i$ in terms of the particle
density (equal to $n_e$) and the polarisation contribution (due to
$\phi$), as
\begin{equation}
n_i = (1+b/2)n_e + \tau_i^{-1}b\phi
\end{equation}
expanding in powers of $b$ and keeping the $O(1)$ and $O(b)$ terms.
This is now converted into configuration space identifying $b$ with
$-\tau_i\ddpp$, leaving
\begin{equation}
n_i = n_e - \ddpp\phi - \half\ddpp p_i
\end{equation}
where $p_i=\tau_i n_e$.  The FLR-corrected potential is given as
\begin{equation}
\phig = (1-b/2)\phi \to \phi + {\tau_i\over 2}\ddpp\phi
\end{equation}
up to $O(b)$.  Here and below, $\ddpp$ is normalised against $\rs^{-2}$,
so that $\rho_i^2\ddpp$ becomes $\tau_i\ddpp$.

Now we use the equations for $n_e$ and $n_i$ to
find the vorticity equation, using these forms to eliminate $n_i$ and
$\phig$ in terms of $n_e$ and $\phi$.
First, the ion density equation becomes
\begin{eqnarray}
& &\ptt{}\LP n_e - \ddpp\phi - \half\ddpp p_i\RP
	+ [\phi,n_e] - [\phi,\ddpp\phi] - \half[\phi,\ddpp p_i]
	+ \half[\ddpp\phi,p_i]
\nonumber\\ & & \hskip 6 cm{}
	+ \dpl\upl = \kappacv(\phi+p_i)
\label{eqned}
\end{eqnarray}
where under $\kappacv$ only the $O(1)$ terms are kept.  The terms under
$\ppt{}$ are from $n_i$.  The first three bracket terms are from
$[\phi,n_i]$, and the last bracket term is from the difference
$\phig-\phi$.  We manipulate the bracket terms involving $\ddpp$ as
follows
\begin{equation}
[\phi,\ddpp\phi]=\div[\phi,\dpp\phi] - [\dpp\phi,\dpp\phi]
\end{equation}
\begin{equation}
[\phi,\ddpp p_i]=\div[\phi,\dpp p_i] - [\dpp\phi,\dpp p_i]
\end{equation}
\begin{equation}
[\ddpp\phi,p_i]=\div[\dpp\phi,p_i] - [\dpp\phi,\dpp p_i]
	=\ddpp[\phi,p_i] - \div[\phi,\dpp p_i]- [\dpp\phi,\dpp p_i]
\end{equation}
so that with cancellations (noting also $[\dpp\phi,\dpp\phi]$ vanishes)
we obtain
\begin{eqnarray}
& &\ptt{}\LP n_e - \ddpp\phi - \half\ddpp p_i\RP
	+ [\phi,n_e] 
	- \div[\phi,\dpp\phi] - \div[\phi,\dpp p_i]
	+ \half\ddpp[\phi,p_i] 
\nonumber\\ & & \hskip 6 cm{}
	+ \dpl\upl = \kappacv(\phi+p_i)
\end{eqnarray}
Then, under $\ddpp$ we replace the bracket $[\phi,p_i]$ with
$-\ppt{p_i}$ noting all the other terms are $O(b)$ corrections to
the $\dpl$ and $\kappacv$ terms, so that
\begin{eqnarray}
& &\ptt{}\LP n_e - \ddpp\phi - \ddpp p_i\RP
	+ [\phi,n_e] 
	- \div[\phi,\dpp\phi] - \div[\phi,\dpp p_i]
\nonumber\\ & & \hskip 6 cm{}
	+ \dpl\upl = \kappacv(\phi+p_i)
\end{eqnarray}
Finally, we combine $\ppt{}$ and $[\phi,]$ into $d/dt$, and $\phi$ and
$p_i$ into $W$, obtaining
\begin{equation}
\dtt{n_e} - \div\dtt{}\dpp W + \dpl\upl = \kappacv(\phi+p_i)
\label{eqnef}
\end{equation}
This is the ion density equation in the isothermal fluid model, wherein
$p_i=\tau_i n_e$ and by quasineutrality the particle (not gyrocenter)
densities are equal.
Subtraction of Eq.\ (\ref{eqnef}) from Eq.\ (\ref{eqne}) in Section
\ref{secfluidenergy}, we find 
\begin{equation}
\div\dtt{}\grad W = \dpl\Jpl - (1+\tau_i)\kappacv(n_e)
\end{equation}
where we have inserted $\upl-\vpl=\Jpl$ and $p_e+p_i=(1+\tau_i)n_e$.
This is the same as Eq.\ (\ref{eqvor}) in Section \ref{secfluidenergy},
under the 
isothermal gyro-Bohm normalised forms $p_e=n_e$ and $p_i=\tau_i n_e$ and
$G=0$. 

The rest of the
isothermal fluid equations are
\begin{equation}
\dtt{n_e} + \dpl(\upl-\Jpl) = \kappacv(\phi-n_e)
\end{equation}
\begin{equation}
\dtt{\upl} + (1+\tau_i)\dpl n_e = 0
\end{equation}
\begin{equation}
\beta_e\ptt{\Apl}+\mu_e\dtt{\Jpl} + \dpl(\phi-n_e) = - 0.51\mu_e\nu_e\Jpl
\end{equation}
and these satisfy the energetics as given in Section
\ref{secfluidenergy}, without the temperature dynamics (the latter
inclusive of $G$ and the heat fluxes).  The resistive 
dissipation includes the $0.51$ coefficient from Ref.\ \cite{Braginskii}
and the collision frequency $\nu_e$ is normalised against $c_s/\Lpp$.
What can be termed ``gyroviscous correspondence'' to the fluid model
is thereby proved for the isothermal case, for any occurrence of
nonlinearity within the local ordering.

\subsection{With temperature dynamics
\label{secthermal}}

Now we return to the version of the model with all the temperature
dynamics, involving $\tipl$ and $\tipp$.  We will make use of the particle
representations of these variables.  For the density the relation
between the particle and gyrocenter representations is given by the
polarisation equation, Eq.\ (\ref{eqpola}), whose thermal version is
(Eq.\ 92 of Ref.\ \cite{GEM}),
\begin{equation}
\Gamma_1 n_i + \Gamma_2 \tipp + {\Gamma_0-1\over\tau_i}\phi = n_e
\label{eqpol}
\end{equation}
The $\Gamma_1$ and $\Gamma_2$ are the gyroaveraging operators.  
Eq.\
(\ref{eqpol}) is the closure 
approximation to the gyrokinetic polarisation equation,
\begin{equation}
\sumsp\int\dW \LB eJ_0(\delta f) + e^2{J_0 F^M J_0-F^M\over T}\phi\RB = 0
\label{eqpolk}
\end{equation}
where $\delta f$ denotes the distribution function,
$\int\dW$ represents integration over velocity space, the sum is
over species, $J_0$ with argument $\kpp\vpp/\Omega$ acting upon $\delta
f$ or $\phi$ is the orbit averaging operator, $F^M$ is the background
Maxwellian, and $e$ and $T$ with $n$ and $M$ are the species constants
giving the charge, background density and temperature, and mass.
The polarisation equation comes from setting the particle charge
density to zero \cite{Lee83}, and the polarisation term itself
ultimately arises from the transformation from gyrocenter to
particle phase space \cite{Hahm88}.  The closure approximation to
$\int\dW F^MJ_0$ is $\Gamma_1$, with argument $b$.  The closure form is
given by $\Gamma_0^{1/2}$ by correspondence to linear kinetic theory
\cite{Dorland93}.  The second operator $\Gamma_2$ is given by the
logarithmic derivative of $\Gamma_1$ with respect to $b$
no matter the form chosen for
$\Gamma_1$, since
\begin{equation}
T\pTT{\Gamma_1} = \int\dW \LP{\mu B\over T}- 1\RP F^MJ_0
\end{equation}
The Pad\'e approximant forms for $\Gamma_1$ and $\Gamma_2$ are
\begin{equation}
\Gamma_1 = {1\over 1+b/2} \qqquad \Gamma_2 = {-b^2/2\over (1+b/2)^2}
\end{equation}
as given in Ref.\ \cite{Beer96}.  Hence, the gyroaveraged potential and
its FLR correction as given in Eq.\ (\ref{eqphig}) result from
\begin{equation}
\phig = \Gamma_1\phi \qqquad \vorg = \Gamma_2\phi
\label{eqvorg}
\end{equation}
That these are the same operators as the ones in Eq.\ (\ref{eqpol}), and
that $\phig$ and $\vorg$ are associated with $n_i$ and $\tipp$, are
fundamentals underlying the free energy conservation of the model
\cite{GEM}. 
The form of the nonlinear terms in the $\tipp$-equation results from the
next higher moment with $\mu B$ and applying free energy conservation as
a constraint.  Further detail on this and FLR closure in general is
given in Refs.\ \cite{GEM,gem2}, which update Refs.\
\cite{Dorland93,Beer96}. 

In these terms the particle (space) representations for the three state
variables for the ions are
\begin{equation}
n_{sp} = \Gamma_1 n_i + \Gamma_2 \tipp + {\Gamma_0-1\over \tau_i}\phi
\end{equation}
\begin{equation}
\tipl{}_{sp} = \Gamma_1 \tipl
\end{equation}
\begin{equation}
\tipp{}_{sp} = \Gamma_1 \tipp + \Gamma_2 (n_i+2\tipp) 
	+ 2{\Gamma_1\Gamma_2\over \tau_i}\phi
\end{equation}
all arising from corresponding moments of Eq.\ (\ref{eqpolk}).  The 
first is the same as the polarisation equation.  In the second, resulting
from the $M\wpl^2-T$ moment, the parallel and perpendicular velocity space
integrals separate, and the $\phi$ piece vanishes.  In the third, the
moment of $\mu B-T$ over the $J_0$ operator gives rise to the same
factor of $(\Gamma_1+2\Gamma_2)$ as in the nonlinearities in the $\tipl$
equation (Eq.\ \ref{eqtipp}) itself, in addition to the term
$\Gamma_1\Gamma_2$ coming from the moment of $\mu B-T$ over the $J_0^2$
operator.

As in the isothermal case, 
the fluid equations are found by constructing the time
derivatives of these variables in the particle (not gyrocenter)
representation.  The low-$\kpp$ limit is taken, with the $O(b)$
corrections kept only in the nonlinear advection terms.  The equations
for $n_i$ and $\tipp$ occur together, ultimately due to the way $J_0$
through its $b$-dependence mixes the perpendicular moments.  Up to
$O(b)$ the particle representations are given by
\begin{equation}
n_i{}_{sp} = n_i + \ddpp\phi+{\tau_i\over 2}\ddpp(n_i+\tipp)
\label{eqnisp}
\end{equation}
\begin{equation}
\tipp{}_{sp} = \tipp + \ddpp\phi+{\tau_i\over 2}\ddpp(n_i+3\tipp)
\label{eqtippsp}
\end{equation}
\begin{equation}
\tipl{}_{sp} = \tipl + {\tau_i\over 2}\ddpp\tipl
\label{eqtiplsp}
\end{equation}
with the terms on the right sides understood to be in the gyrocenter
representation. 
In the $O(b)$ terms the representations are equivalently the particle or
gyrocenter ones, so that the inverses of Eqs.\
(\ref{eqnisp}--\ref{eqtiplsp}) are given by
\begin{equation}
n_i{}_{gy} = n_i - \ddpp\phi-{\tau_i\over 2}\ddpp(n_i+\tipp)
\label{eqnigy}
\end{equation}
\begin{equation}
\tipp{}_{gy} = \tipp - \ddpp\phi-{\tau_i\over 2}\ddpp(n_i+3\tipp)
\label{eqtippgy}
\end{equation}
\begin{equation}
\tipl{}_{gy} = \tipl - {\tau_i\over 2}\ddpp\tipl
\label{eqtiplgy}
\end{equation}
with the terms on the right sides understood to be in the particle
representation. 
The gyroreduced potentials in Eq.\ (\ref{eqvorg}) by
\begin{equation}
\phig = \phi+{\tau_i\over 2}\ddpp\phi \qqquad
	\vorg = {\tau_i\over 2}\ddpp\phi
\end{equation}
The partial time derivatives of Eqs.\ (\ref{eqnigy}--\ref{eqtiplgy}) are
taken, and then Eqs.\ (\ref{eqni}--\ref{eqtipl}) are used to evaluate
the right hand sides, as was done in Eq.\ (\ref{eqned}) above.

For the density the result of the substitution is 
\begin{eqnarray}
& &\ptt{}\LP n_i - \ddpp\phi - \half\ddpp p_i\RP
	+ [\phi,n_e] - [\phi,\ddpp\phi] - \half[\phi,\ddpp p_i]
	+ \half[\ddpp\phi,p_i]
\nonumber\\ & & \hskip 6 cm{}
	+ \dpl\upl = \kappacv\LP\phi+{\pipl+\pipp\over 2}\RP
\end{eqnarray}
where $\ddpp p_i=\tau_i\ddpp(n_i+\tipp)$, linearised as before,
represents the combining of the $\ddpp n_i$ and $\ddpp\tipp$ terms.  The
manipulations of the $\ddpp$ operators are done exactly as before, and
the result is 
\begin{equation}
\dtt{n_i} - \div\dtt{}\dpp W + \dpl\upl 
	= \kappacv\LP\phi+{\pipl+\pipp\over 2}\RP
\label{eqnifa}
\end{equation}
For the perpendicular temperature the result of the substitution is 
\begin{eqnarray}
& &\ptt{}\LP \tipp - \ddpp\phi - \half\ddpp p_i - \tau_i\ddpp\tipp\RP
\nonumber\\ & & \hskip 1 cm{}
	+ [\phi,\tipp] - [\phi,\ddpp\phi] - \half[\phi,\ddpp p_i]
		- [\phi,\ddpp\tau_i\tipp]
	+ \half[\ddpp\phi,p_i] + [\ddpp\phi,\tau_i\tipp]
\nonumber\\ & & \hskip 3 cm{}
	+ \dpl\qipppl = \kappacv\LP{\phi+\pipp+3\tau_i\tipp\over 2}\RP
	+2\nu_i G
\end{eqnarray}
where FLR corrections to the $\kappacv$ terms are dropped as before.
The nonlinear terms proportional to $\tipp$ arise from the extra factors
of $2\tipp$ in Eqs.\ (\ref{eqtipp},\ref{eqtippgy}).
The
manipulations of the $\ddpp$ operators are done exactly as before, and
the result is 
\begin{equation}
\dtt{\tipp} - \div\dtt{}\dpp W - 2\tau_i\div\dtt{}\dpp\tipp
	+ \dpl\qipppl = \kappacv\LP{\phi+\pipp+3\tau_i\tipp\over 2}\RP
	+2\nu_i G
\label{eqtippfa}
\end{equation}
with the last of the nonlinear time derivative terms representing the
perpendicular part
of the polarisation heat flux.
For the parallel temperature the result of the substitution is 
\begin{eqnarray}
& &\ptt{}\LP \tipl - {\tau_i\over 2}\ddpp\tipl\RP
	+ [\phi,\tipl] - {\tau_i\over 2}[\phi,\ddpp\tipl]
	+ {\tau_i\over 2}[\ddpp\phi,\tipl]
\nonumber\\ & & \hskip 4 cm{}
	+ 2\dpl(\upl+\qiplpl) = \kappacv\LP\phi+\pipl+2\tau_i\tipl\RP
	-4\nu_i G
\end{eqnarray}
where FLR corrections to the $\kappacv$ terms are dropped as before.
The nonlinear terms proportional to $\tipl$ arise from the extra factors
of $\tipl$ in Eqs.\ (\ref{eqtipl},\ref{eqtiplgy}).
The
manipulations of the $\ddpp$ operators are done exactly as before, and
the result is 
\begin{equation}
\half\dtt{\tipl} - {\tau_i\over 2}\div\dtt{}\dpp\tipl
	+ \dpl(\upl+\qiplpl) = \kappacv\LP{\phi+\pipl+2\tau_i\tipl\over 2}\RP
	-2\nu_i G
\label{eqtiplfa}
\end{equation}
with the last of the nonlinear time derivative terms representing the
parallel part of the polarisation heat flux.
The two temperature equations (Eqs.\ \ref{eqtippfa},\ref{eqtiplfa}) are
added to provide the final temperature equation
\begin{equation}
\threehalves\dtt{T_i} - \div\dtt{}\dpp W 
	- \fivehalves\tau_i\div\dtt{}\dpp T_i
	+ \dpl(\upl+\qipl) = \kappacv\LP\phi+p_i
	+\fivehalves\tau_i T_i + 2G\RP
\label{eqtif}
\end{equation}
where we use $T_i=(2\tipp+\tipl)/3$ and
$G=\tau_i(\tipl-\tipp)/6$ and
$\qipl=\qipppl+\qiplpl$.
The anisotropy dissipation term cancels.  The first factor of $G$ is
from the pressures (diamagnetic flow in the fluid model), and the second
is from the temperatures (diamagnetic heat fluxes) and is neglected in
the fluid model.  In terms of $p_i$ and $G$ the density equation
(Eq.\ \ref{eqnifa}) becomes
\begin{equation}
\dtt{n_i} - \div\dtt{}\dpp W + \dpl\upl = \kappacv\LP\phi+p_i+G\RP
\label{eqnif}
\end{equation}
Subtraction of Eq.\ (\ref{eqnif}) from Eq.\ (\ref{eqne}) recovers Eq.\
(\ref{eqvor}) above, and then addition of Eq.\ (\ref{eqvor}) to Eq.\
(\ref{eqtif}) recovers Eq.\ (\ref{eqti}) above (except for the second
factor of $G$ which the fluid model doesn't keep), and the
correspondence in the nonlinear polarisation terms is thereby proved.

\section{Gyrofluid temperature anisotropy and fluid parallel viscosity}

We now turn to the less obscure parts of the correspondence between
gyrofluid and fluid equations.  
The
general pressure tensor arises from
\begin{equation}
\vec P = \int\dW m\vec w\vec w\,f(\vec w)
\end{equation}
where the integration is over velocity space and $\vec w$ is the random
kinetic velocity in the co-moving frame with fluid velocity $\vec u$.
In the Braginskii fluid equations $f$ is assumed to be a Maxwellian $f_0$
with
variable density $n$ and temperature $T$, and also flow $\vec u$, where
all of $n$, $T$, $\vec u$ are arbitrarily variable, i.e., including all
dynamics as well as the background (reduction to low frequency equations
under drift ordering comes later).  Then, the corrections to $f_0$ are
considered to be of the form $f_1=\Phi(\vec w) f_0$, and $\Phi$ solved
for in terms of Sonine polynomials, with arbitrarily large collision
frequency and small gyroradius, as well as small mean free path (i.e., all
gradients are assumed to represent small corrections to local
thermodynamic equilibrium, LTE).  Besides
the specific conductive heat flux ($\vec q/nT$), the temperature 
anisotropy is assumed to be small.  Hence $\vec P$ is split in terms of
an isotropic part and a trace-free correction,
\begin{equation}
\vec P = p\,\vec g + \Pi
\end{equation}
where $\vec g$ is the metric tensor.  The diagonal elements of $\Pi$
represent the parallel viscosity.  These pressure contributions may be
written, separately from any non-diagonal $\Pi$ contributions, as
\begin{equation}
\vec P = \ppp\,\vec g + (\ppl-\ppp)\bunit\bunit
\end{equation}
where $\bunit=\vec B/B$ is the magnetic unit vector, hence
\begin{equation}
\Pi = \deltap\bunit\bunit - \third\deltap\,\vec g
\end{equation}
with isotropic pressure and deviation given by
\begin{equation}
p = {2\ppp+\ppl\over 3} \qqquad \deltap = \ppl-\ppp
\end{equation}
We also have
\begin{equation}
\ppl = p + \twothirds\deltap \qqquad \ppp = p - \third\deltap
\end{equation}
to assist the evaluation of gyrofluid combinations.

It is important to note that with the inhomogeneous magnetic field
$\div(\bunit\bunit)$ contributes to the general divergence $\div\vec
P$.  Then, in the reduction to low frequency we obtain
\begin{equation}
\bunit\cdot(\div\vec P) = \dpl\ppl - \deltap\dpl\log B
\end{equation}
for the parallel pressure force, which is covered by the parallel
gradient and magnetic pumping terms in the gyrofluid moment equation for
the parallel velocity.  In the perpendicular drifts we make the same
approximations as in the gyrofluid model itself:
\begin{equation}
\div\drift\grad p \to -2\grad\log B\cdot\drift\grad p
\qqquad
\bdel\bunit \to \grad\log B
\end{equation}
and hence also
\begin{equation}
\div\drift(\bdel\bunit) \to -2\grad\log B\cdot\drift\grad\log B = 0
\end{equation}
We can then find
\begin{equation}
\div\drift(\div\vec P) \to -\grad\log B\cdot\drift\grad(\ppl+\ppp)
\end{equation}
which is covered by the same curvature terms in the gyrofluid moment
equation for the density.
The gyrofluid model then also includes FLR effects,
wherein $\dpl\phi$ becomes $\dpl\phig$ and under derivatives and in the
magnetic pumping terms $\tau_i\tipp$ becomes $\tau_i\tipp+\vorg$.

Hence, the pressure/temperature anisotropy enters the continuity
equations for ions the same way as in the fluid model, although via a
different route: grad-B and curvature drifts for gyrocenters
rather than the pressure tensor for particles.  It remains to obtain the
anisotropy itself.  The gyrofluid equations for $\tipp$ and $\tipl$ for
the ion species are given by
\begin{eqnarray}
& &\half\ptt{\tipl}+\half[\phig,\tipl] + B\dpl{\upl+\qiplpl\over B}
	- (\upl+\qipppl)\dpl\log B 
\nonumber\\ & &\qquad{}
	= \kappacv\LP{\phig+\tau_i n_i+3\tau_i \tipl\over 2}\RP
	- {\nu_i\over 3\pi_i}[\tau_i(\tipl-\tipp)-\vorg]
\label{eqtiplv}
\end{eqnarray}
\begin{eqnarray}
& &\ptt{\tipp}+[\phig,\tipp]+[\vorg,(n_i+2\tipp)]
	+ B\dpl{\qipppl\over B}
	+ (\upl+\qipppl)\dpl\log B 
\nonumber\\ & &\qquad{}
	= \kappacv\LP{\phig+\tau_i n_i+4\tau_i \tipp+4\vorg\over 2}\RP
	+ {\nu_i\over 3\pi_i}[\tau_i(\tipl-\tipp)-\vorg]
\label{eqtippv}
\end{eqnarray}
where the $\dpl\log B$ terms give the magnetic pumping of anisotropy and
the $\nu_i$ terms its collisional dissipation, and $\pi_i$ is a
numerical constant which we eventually adjust to obtain correspondence.
If we add these equations the total temperature equation results, and
the magnetic pumping and dissipation cancel, leaving
\begin{equation}
\threehalves\dtt{T_i} 
	+ \hbox{(FLR)}
	+ B\dpl{\upl+\qipl\over B}
	= \kappacv\LP\phi+\tau_i n_i+{7\over 2}\tau_i T_i 
	+ \third\tau_i\deltat\RP
\label{eqtiv}
\end{equation}
where ``FLR'' denotes the FLR corrections which eventually become the
polarisation terms as established above and $T_i=(2\tipp+\tipl)/3$ and
$\qipl=\qiplpl+\qipppl$ as before.  The anisotropy is
$\deltat=\tipl-\tipp$.  If we instead subtract Eq.\ (\ref{eqtippv}) from
twice Eq.\ (\ref{eqtiplv}) forming an equation for $\deltat$, we find
\begin{eqnarray}
& &\ptt{\deltat} + [\phig,\deltat] - [\vorg,(n_i+2\tipp)]
	+ B\dpl{2\qiplpl-\qipppl\over B}
\nonumber\\ & &\qquad\hskip 1 cm{}
	- 3(\upl+\qipppl)\dpl\log B 
	- \kappacv\LP{\phig-\phi+\vorg\over 2}-{8\over 3}\tau_i\deltat\RP
	- {\nu_i\over \pi_i}\vorg
\nonumber\\ & &\qquad{}
	= \half\kappacv\LP\phi + \tau_i n_i + \tau_i T_i\RP
	- 2B\dpl{\upl\over B} - {\nu_i\over \pi_i}\tau_i\deltat
\label{eqdtv}
\end{eqnarray}
having arranged terms such that the collisional and velocity divergence
terms are on the right side and the nonlinearities, magnetic pumping,
FLR, anisotropy in curvature terms,
and heat flux effects are on the left side.  The Braginskii assumptions
are essentially that the left side terms are small, even though that is
obviously not the case in the curvature terms, as pointed out before
\cite{Smolyakov97}. 

If the Braginskii assumptions are taken, then we find
\begin{equation}
\tau_i\deltat = {\pi_i\over 2\nu_i}\LB
	\kappacv\LP\phi + \tau_i n_i + \tau_i T_i\RP
	- 4B\dpl{\upl\over B}\RB
\end{equation}
which is equivalent to 
\begin{equation}
G = {\pi_i\over 12\nu_i}\LB(\grad\log B^2\cdot\uperp)
	-4(\div\upl\bunit)\RB
\end{equation}
This is
the form given by Ref.\ \cite{Rogers98}, and correspondence is thereby proved.
It is important to note, however, that this regime is never reached even
in deep edge turbulence.  Tokamak edge turbulence typically has $\nu_i$
about two orders of magnitude slower than nonlinear advection.  Worse
than this, the nonlinear advection is the largest effect in Eq.\
(\ref{eqdtv}), larger than any of $c_s/R$ or $c_s/qR$ or $\nu_i$, even
for zonal flows.  Hence the dissipation of ion flows cannot be properly 
modelled by $\kappacv(G)$ in the vorticity equation or $\dpl G$ in the
parallel velocity equation, with either $\nu_i$ or $c_s/qR$ or $c_s/R$ 
as the controlling frequency.
These forms will overestimate the
dissipation effects and will enforce a particular phase shift between
viscosity and the variables determining the ion flow ($\phi$, $n_i$,
$T_i$, and $\upl$).  Use of the collisional form with $\nu_i$ will very
strongly overestimate ion flow damping in tokamak edge regimes,
ultimately corrupting any investigation of bifurcation dynamics.
Nevertheless, we set $\pi_i=0.96$ for ions and $\pi_e=0.73$ for
electrons to obtain correspondence to the Braginskii regime in the
gyrofluid equations.

\section{Gyrofluid heat fluxes and the collisional fluid limit}

The simplest correspondence is in the heat flux equations.  In the
gyrofluid model the parallel heat fluxes (parallel and perpendicular
energy components) are dynamical variables with their own equations.
However, the Braginskii limit assumes that all of the time scales
involved in advection, divergences, dissipation, etc., of the heat
fluxes are slow, with the exception of collisional dissipation.  The
dissipation balances the forcing represented by the temperature
gradient.  
When the parallel heat flux components (Eqs.\ 103,104 of Ref.\ 23)
are added to form the
equation for $\qipl=\qiplpl+\qipppl$ the anisotropy dissipation
effects cancel, leaving
\begin{eqnarray}
& &\mu_i\dtt{\qipl}+\hbox{(FLR)}+\hbox{(LD)}
	-\mu_i\tau_i\kappacv(2\upl+4\qiplpl+3\qipppl)
	\nonumber\\ & &\qquad{}
	= -\tau_i\dpl\LP\threehalves\tipl+\tipp\RP
	-{5/2\over\kappa_i}\mu_i\tau_i\nu_i\qipl
\end{eqnarray}
where $\kappa_i$ is the thermal conduction coefficient, and 
``FLR'' denotes the FLR corrections (including the appearance of
$\vorg$ under $\dpl$)
and ``LD'' the Landau damping
dissipation.  All of the terms on the left side scale with advection or
are slower.  The terms on the right side are the ones left after the
Braginskii ordering is taken --- assuming that $\nu_i$ overpowers
advection.  Neglecting the temperature anisotropy the factors of $5/2$
cancel and we have
\begin{equation}
\qipl = -{\kappa_i\over\mu_i\nu_i}\dpl T_i
\end{equation}
which is the Braginskii formula.  We set $\kappa_i=3.9$ to set the
quantitative correspondence.

For electrons we additionally have the mixing of the moments under the
collisional dissipation.  The re-expression of the thermal force as such
was given in Ref.\ \cite{dalfloc}, whose two salient equations in the
fluid model are
\begin{equation}
\beta_e\ptt{\Apl}+\mu_e\dtt{\Jpl}=\dpl(n_e+T_e-\phi)
	-\mu_e\nu_e\LB \eta\Jpl+{\alpha_e\over\kappa_e}(\qepl+\alpha_e\Jpl)\RB
\label{eqdalfjpl}
\end{equation}
\begin{equation}
\mu_e\dtt{\qepl}+\hbox{(LD)}=-\fivehalves\dpl T_e
	-{5/2\over\kappa_e}\mu_e\nu_e(\qepl+\alpha_e\Jpl)
\label{eqdalfqe}
\end{equation}
where the coefficients $\eta$, $\kappa_e$, $\alpha_e$ are for
resistivity, thermal conduction, and the thermal force, respectively.
If the nonlinear advection and Landau damping are assumed small in 
Eq.\ (\ref{eqdalfqe}) then the Braginskii formula
\begin{equation}
\qepl+\alpha_e\Jpl = -{\kappa_e\over\mu_e\nu_e}\dpl T_e
\label{eqbragqe}
\end{equation}
with $\kappa_e=3.2$ and $\alpha_e=0.71$ for pure hydrogen, is
recovered.  Then, insertion of this into Eq.\ (\ref{eqdalfjpl}) 
gives
\begin{equation}
\beta_e\ptt{\Apl}+\mu_e\dtt{\Jpl}=\dpl(n_e+T_e-\phi)+\alpha_e\dpl T_e
	-\eta\mu_e\nu_e\Jpl
\label{eqbragjpl}
\end{equation}
which recovers the Braginskii Ohm's law if $\eta=0.51$ is chosen.

This dissipation model was built into the electron gyrofluid moment
equations in order to obtain this correspondence, in both Refs.\
\cite{aps99,GEM}.  The electron heat flux equations are
\begin{eqnarray}
& &\mu_e\ptt{\qeplpl} + \mu_e\alde\qeplpl + \mu_e[\phi_e,\qeplpl]
	= - \threehalves\dpl\tepl
	- \mu_e\kappacv\LP{3\vpl+8\qeplpl\over 2}\RP
	\nonumber\\ & &\qquad{}
	- {(5/2)\over\kappa_e}\mu_e\nu_e
		\LP\qeplpl+0.6\alpha_e\Jpl\RP 
		+ 1.28\nu_e\LP\qeplpl-1.5\qepppl\RP
\end{eqnarray}
\begin{eqnarray}
& &\mu_e\ptt{\qepppl} + \mu_e\alde\qepppl + \mu_e[\phi_e,\qepppl]
	+ \mu_e[\vor_e,(\vpl+2\qepppl)]
	\nonumber\\ & &\qquad{}
	= - \dpl(\tepp-\vor_e)
	- \mu_e\kappacv\LP{\vpl+6\qepppl\over 2}\RP
	- \LP\tepp-\tepl-\vor_e\RP\dpl\log B
	\nonumber\\ & &\qquad{}
	- {(5/2)\over\kappa_e}\mu_e\nu_e
		\LP\qepppl+0.4\alpha_e\Jpl\RP 
		- 1.28\nu_e\LP\qeplpl-1.5\qepppl\RP
\end{eqnarray}
from Eqs.\ (103,104) with additions in Eqs.\ (114,118,119)
of Ref.\ \cite{GEM}.  
Electron FLR corrections are
kept, with $\phi_e$ and $\vor_e$ the corresponding potentials.  Adding
these to form the total, neglecting FLR effects, magnetic
pumping, and curvature terms, we find
\begin{equation}
\mu_e\dtt{\qepl} + \hbox{(LD)} 
	= - \dpl\LP\threehalves\tepl+\tepp\RP
	- {(5/2)\over\kappa_e}\mu_e\nu_e(\qepl+\alpha_e\Jpl)
\end{equation}
noting the anisotropy dissipation terms cancel.  Now assuming $\nu_e$
overcomes nonlinear advection or Landau damping, and neglecting
$\deltat$, we find 
\begin{equation}
\qepl+\alpha_e\Jpl = -{\kappa_e\over\mu_e\nu_e}\dpl T_e
\end{equation}
which is the same as Eq.\ (\ref{eqbragqe}) above, i.e., the Braginskii
formula.

The electron gyrofluid parallel velocity equation is
\begin{eqnarray}
& &\beta_e\ptt{\Apl}-\mu_e\ptt{\vpl}-\mu_e[\phi_e,\vpl]-\mu_e[\vor_e,\qepppl]
	\nonumber\\ & &\qquad{}
	= -\dpl\LP\phi_e-n_e-\tepl\RP
	- \mu_e\kappacv\LP{4\vpl+2\qeplpl+\qepppl\over 2}\RP
	\nonumber\\ & &\qquad{}
	-\LP\vor_e-\tepp+\tepl\RP\dpl\log B
	-\mu_e\nu_e\LB \eta\Jpl+{\alpha_e\over\kappa_e}(\qeplpl+\qepppl+\alpha_e\Jpl)\RB
\end{eqnarray}
from Eq.\ (100) with additions in Eq.\ (115) of Ref.\ \cite{GEM}.
Neglecting FLR effects, magnetic pumping, curvature terms, adding
$\qeplpl+\qepppl=\qepl$, replacing $\tepl$ by $T_e$ (neglecting $\deltat$)
and setting $\vpl=-\Jpl$ (effectively
neglecting finite $\mu_e$ corrections), we find
\begin{equation}
\beta_e\ptt{\Apl}+\mu_e\dtt{\Jpl}=\dpl(n_e+T_e-\phi)
	-\mu_e\nu_e\LB \eta\Jpl+{\alpha_e\over\kappa_e}(\qepl+\alpha_e\Jpl)\RB
\end{equation}
which is the same as Eq.\ (\ref{eqdalfjpl}) above.  Then going to the
Braginskii limit by inserting $\qepl$ from Eq.\ (\ref{eqbragqe}), we
find Eq.\ (\ref{eqbragjpl}), which is the Braginskii version.  

It has been pointed out that curvature terms should appear in the fluid
model's equations for $\Jpl$ and $\upl$ \cite{Smolyakov97}.  This is
indeed the case, but in that event the involved terms are the same as
the ones in the gyrofluid model.  Correspondence in the heat fluxes and
the Ohm's law is thereby proved.

\section{Conclusions}

This work has considered the low frequency Braginskii fluid drift
equations on the one hand and the 
electromagnetic, transcollisional gyrofluid equations 
on
the other.  In both cases the model is comprehensive enough to treat
temperature dynamics of both species and has a free energy functional
which is conserved in the absence of dissipation and external drive.  By
considering the entirety of the Braginskii limit --- collision frequency
larger than advection or transit frequency, specific heat flux smaller
than fluid velocity, and small gyroradius --- the sets of equations and
each one's conserved free energy have been shown to be one and the
same.  Whether or not the Braginskii limit is ever reached is a separate
question.  But the low frequency Braginskii fluid drift
equations have been shown by correspondence to be a fully contained
subset of the electromagnetic, transcollisional gyrofluid equations.

One substantial advantage of the gyrofluid model is the fact that in all
the terms involving derivatives, only scalar quantities are involved.
Instead of vector or tensor components, it has gyroreduced potentials or
charge densities which involve Hermitian operators (enabling the free
energy conservation).  This leads to numerical schemes which are easier
to formulate.  All the nonlinear terms have the Poisson bracket
structure, for which the Arakawa spatial discretisation scheme is
uniquely suited \cite{Arakawa}.  Coupled with a timestep that is highly
accurate, requires only one evaluation of the terms per step, and is
stable for waves \cite{Karniadakis}, we have the best scheme found so
far for this type of microturbulence in magnetised plasmas \cite{TYR}.
Its use for the gyrofluid equations is detailed in Ref.\ \cite{GEM}.
With these advantages together with the correspondence to the Braginskii
fluid drift equations, some of the mystery surrounding the efficacy or
validity of the gyrofluid model for tokamak edge turbulence should be
alleviated.

\par\vfill\eject

{
\bibliography{../paper}
\bibliographystyle{aip}
}

\par\vfill\eject

\end{document}